\documentclass[reprint,superscriptaddress,amsmath,amssymb,aps,pra,floatfix]{revtex4-1}
\usepackage{epsfig}
\usepackage[caption=false]{subfig}
\usepackage{graphicx}
\usepackage{dcolumn}
\usepackage{bm}
\usepackage{xcolor}

\begin{document}


\title{\textit{First-priniciple} based study of transport properties of non-trivial topological fermions of CoSi
}

\author{Paromita Dutta}
 \altaffiliation{dutta.paromita1@gmail.com}
 \affiliation{%
School of Basic Sciences, Indian Institute of Technology Mandi, Kamand, Himachal Pradesh-175075, India}%

\author{Sudhir K. Pandey}
\altaffiliation{sudhir@iitmandi.ac.in} 
\affiliation{%
School of Engineering, Indian Institute of Technology Mandi, Kamand, Himachal Pradesh-175075, India
}%

\date{\today}

\begin{abstract}

Recently, CoSi has been identified to have unconventional electronic topology due to lack of inversion center in its B20 cubic structure. The electronic topology has been reported to be present at three nodal points found in the band structure. Two of these nodal points are situated at the $\Gamma$ (G1 \& G2) and one at $R$ (R1) point. Based on this, we present a study where various transport coefficients are investigated by using \textit{first-principle} based DFT method for the temperature (T) range 40-300 K. For the chemical potential ($\mu$) corresponding to energies of these nodal points and at the Fermi level (E$_F$), 3D constant energy surfaces are constructed. They have shown that the number of states available at energies of these nodal points and the E$_F$ follows an increasing trend as R1 $>$ G2 $>$ E$_F$ $>$ G1 at T = 0 K. Similar increasing behavior seems to follow by other transport coefficients at different $\mu$ with T rise such as electrical conductivity($\sigma$)/relaxation time($\tau$) ratio and electronic thermal conductivity ($\kappa_e/\tau$ = $\kappa^0$). For example, at T = 100 K, $\sigma / \tau$ $\sim$ 0.13$\times$10$^{20}$ $\Omega^{-1} m^{-1} s^{-1}$ at G1 whereas its value reaches $ \sim$ 0.18$\times$10$^{20}$ $\Omega^{-1} m^{-1} s^{-1}$ at R1 nodal point. However, Seebeck coefficient (\textit{S}) seems to follow the trend as G2 $>$ E$_F$ $>$ R1 $>$ G1 at any given T. The values of \textit{S} are obtained to be positive at the $\mu$ corresponding to the G2, R1 \& E$_F$ (except G1) which is increasing with the rise in T. Also, the dominant charge carriers at G1 point are found to be electrons for T $<$ 225 K whereas for T $>$ 225 K, the charge carriers are obtained to be dominated by holes. Furthermore, the doping concentrations have also been calculated for G1 (electron doping $\sim$ 2.25$\times$10$^{22}$ cm$^{-3}$), G2 \& R1 points (hole doping $\sim$ 4.10$\times$10$^{20}$ cm$^{-3}$ \& 7.22$\times$10$^{22}$ cm$^{-3}$, respectively).

\end{abstract}

\maketitle


\section{Introduction}
\small
 Since decades the family of transition-metal (TM) silicides have attracted a considerable focus due to their obvious feature as the materials for thermoelectric-energy converters \cite{Lange_1997,Fedrov_1995}. The members of this family are potentially cheap and their vast abundance makes them good for thermoelectric applications. Amongst them, CoSi has been found to be one of the candidates for advanced thermoelectric applications \cite{Asanabe_1964, Kim_2002,Nikitin_1970,Alekseeva_1981,Liu_2012,Lue_2004,Sakai_2007}. CoSi  has been reported as semi-metal due to the slight overlap of conduction bands (CB) and the valence bands (VB) at the Fermi level \cite{Rowe_1995,Imai_2001,Pan_2007}. At 300 K, the material has a moderate Seebeck coefficient (\textit{S}) of $\sim$ -80 $\mu$V/K and thermal conductivity ($\boldsymbol \kappa$) of $\sim$ 20 Wm$^{-1}$K$^{-1}$ \cite{Fedrov_1995,Asanabe_1964,Lue_2004}. It has been also reported to be diamagnetic, having a temperature (T) independent susceptibility ($\chi\sim -0.44\times10^{-6}$ emu/g) \cite{Benoit_1955,Wertheim_1966,Wertheim_1972}. However, Petrova \textit{et al.} and Amamou \textit{et al.} have reported that it is diamagnetic over a wide range of T but it shows a paramagnetic behaviour at lower values of T \cite{Petrova_2010,Amamou_1972}.    

CoSi exhibits B20 type structure having a cubic unit cell without an inversion center. It is well known that the absence of inversion center suggests the existence of topological non-trivial electronic states in these types of materials. In light of this, CoSi has been reported to be a candidate of the family of Weyl semi-metals\cite{Ishii_2014,Wehling_2014,Huang_2016}. However, few of the recent density functional theory (DFT) based calculations and experimental studies have shown that CoSi along with other transition-metal monosilicides such as RhSi belongs to the class of topologically nontrivial materials possessing different kinds of fermions other than Weyl \cite{Pauling_1948,Bradlyn_2016,Tang_2017,Chang_2017,Pshenay_2018,Burkov_2017,Burkov_2019,Sanchez_2019,Takane_2019,Rao_2019,Dutta_2021}. In this class of materials, band crossings are found to exist in their electronic band structures. In the vicinity of these band-crossing points, few of the CB and VB touch each other and have linear energy dispersions. These points are often referred to as nodes or nodal points. For instance, in the band structure calculations of CoSi, when performed with spin-orbit coupling, three types of unconventional linear nodal points have been reported. Among the three nodes, two are at the $\Gamma$ point with one having 4-fold degeneracy while the other having 2-fold degeneracy. The third node point is found at $R$ point having 6-fold degeneracy. Thus, at these points, the unconventional charge carriers seem to be found. Considering this aspect, if chemical potential ($\mu$) is positioned close to the energies of these nodal points then there is a possibility of capturing the contribution of the unconventional charge carriers to electronic transport of the material. One of the way through which this can be achieved is by doping of the material. As being one of the prime candidates for advanced thermoelectric applications, many investigations have already been done in order to improve its thermoelectric power, mostly by doping \cite{Lue_2004,Li_2005,Ren_2005,Pan_2007,Longhin_2017,Sun_2013,Kuo_2005,Jian_2020}. It has been reported in these studies that transport coefficients of CoSi could be explained by using semi-metallic band structure model \cite{Fedrov_1995,Asanabe_1964}. Parabolic band dispersions have been presumed  where effective masses of electrons and holes are taken as $\sim$ 2m$_e$ and 4m$_e$-6m$_e$, respectively (here m$_e$ is free electron mass) \cite{Alekseeva_1981,Asanabe_1964,Asanabe_1965,Nikitin_1970,Pan_2007}. Thus, there is an urge of exploring the transport behaviour of these new electronic states as discovered to exist in CoSi at the nodal points. Accordingly, the present study is based on studying the transport behaviour of these nodal points separately by relating them with their constant energy surfaces. 

Based on this, we have investigated the transport coefficients of CoSi by carrying out first principle based DFT calculations at the nodal points (when $\mu$ is at G1, G2 \& R1). The range of T is chosen from 40 - 300 K. Here, G1 \& G2 corresponds to two nodal points found at the $\Gamma$ point and, R1 corresponds to the nodal point found at $R$ point in the band structure. Now, one can also think of correlations which may exist due to the fact that CoSi being one of the members of 3\textit{d} TM. But recently, Dutta \textit{et al.} have  shown that the bands in the vicinity of the Fermi level (E$_F$) are less affected by on-site Coulomb interactions when studied by using dynamical mean field theory method. In-fact, they were similar to the DFT bands \cite{Dutta_2018,Dutta_2019,Dutta_2021}. Thus, for the present study, we have chosen DFT method. Further, using the DFT results, the transport coefficients are studied by using semi-classical theory. To understand the transport behaviours corresponding to these nodal points, the constant energy surfaces for the $\mu$ corresponding to these node points have been constructed. In the analysis of these surfaces, it has been found that the number of states available at energies of the nodal points and the E$_F$ follows an increasing trend as R1 $>$ G2 $>$ E$_F$ $>$ G1. Furthermore, it has been found that the electrical conductivity per unit relaxation time ($\boldsymbol \sigma/\tau$) is maximum at the $\mu$ corresponding to R1 point and least at the $\mu$ corresponding to G1 point, at any given T. It follows an increasing trend as R1 $>$ G2 $>$ E$_F$ $>$ G1 at any given T. For instance, at T = 100 K, $\boldsymbol \sigma/\tau$ $\sim$ 0.13 x 10$^{20}$ $\Omega^{-1} m^{-1} s^{-1}$ at the $\mu$ corresponding to G1 whereas its value reaches $ \sim$ 0.18 x 10$^{20}$ $\Omega^{-1} m^{-1} s^{-1}$ at the $\mu$ corresponding to R1 nodal point. At T = 300 K, the value of $\kappa^0 is \sim$ 2.4 Wm$^{-1}$K$^{-1}$ at E$_F$, which is quite close to the recent experimental value of $\sim$ 3 Wm$^{-1}$K$^{-1}$. However, the Seebeck coefficient (\textit{S}) is seen to follow a different trend as G2 $>$ E$_F$ $>$ R1 $>$ G1. It has shown a positive value corresponding to $\mu$ of G2, E$_F$ and R1 point. Unlike other nodal points, the value of $S$ corresponding to $\mu$ at G1 point is found to be negative for T $<$ 225 K and beyond 225 K, it changes to positive values. In the end, we have also provided the doping concentrations needed for positioning the Fermilevel to the chemical potential of the energies corresponding to these nodal points.  

\section{Computational Details}
\small
In this work, first-principle based DFT calculations have been performed using WIEN2k program \cite{Blaha_2001}. The program is based on full-potential linearized augmented plane wave method. This program is used for the calculation of electronic structure of CoSi in the presence of spin-orbit coupling (SOC). The lattice parameters are taken from the literature \cite{Boren_1933}. PBEsol is chosen as exchange-correlation functional for the calculations \cite{Perdew_2008}. The muffin-tin sphere radii for Co and Si sites are taken as 2.18 and 1.84 Bohr, respectively. The energy convergence limit for the self-consistent cycle is used as 10$^{-4}$ Ryd. Apart form this, the calculation of the temperature dependent transport coefficients have been also performed using the BoltzTraP program \cite{Madsen_2006}. This program is interfaced with WIEN2k package. The transport coeffecients are highly sensitive to the band structure. Hence, a dense k-mesh of size 70 x 70 x 70 has been taken for the proper calculation of transport properties of the material.  

\section{Results and Discussion}

\begin{figure}[tbh]
  \begin{center}
   \includegraphics[width=2.4in]{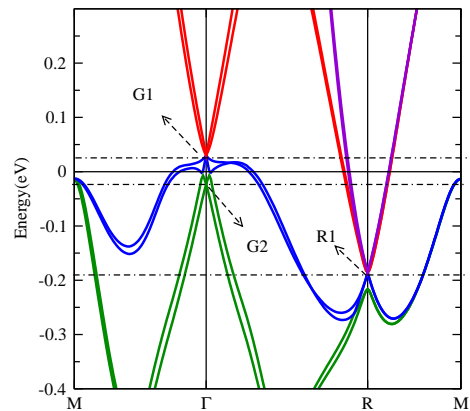}
  \end{center}

  \caption{\small{(colour online) Electronic band structure of CoSi with spin-orbit coupling along the high symmetric directions. Zero energy (solid line) corresponds to the Fermi energy (E$_F$). Dashed line arrows are denoting few specific points corresponding to different values of $\mu$ (dashed lines).}}
 
\end{figure}

In order to understand the transport behaviours of any material, the major inputs are taken from the electronic bands of the crystal. Accordingly, the band structure of CoSi is plotted in Fig. 1 along high symmetric directions, $M-\Gamma-R-M$. Here, it is important to note that the transport properties are discussed corresponding to the four different chemical potentials ($\mu$) where one of the case is when $\mu$ is zero \textit{i.e.}, the Fermi energy (E$_F$). The other three are the energies corresponding to the specific $k$-points (nodal points) where new fermions have been reported for CoSi. These points are situated at the different energies of the $\Gamma$ and \textit{R} points \cite{Tang_2017,Takane_2019}. It has been reported that these nodal points are associated with new nontrivial topological fermions. For instance, 4-fold degenerate bands and 2-fold degenerate bands at $\Gamma$ point are associated with spin-3/2 RSW fermion and spin-1/2 Weyl fermion, respectively. Furthermore, 6-fold degenerate bands at \textit{R} point are associated with double spin-1 fermion. Accordingly, the two nodal points at $\Gamma$ are represented as G1 ($\mu \sim$ 30 meV) and G2 ($\mu \sim$ -23 meV) while the nodal point at \textit{R} point is represented as R1 ($\mu \sim$ -186 meV) also marked in the Fig. 1.

\begin{figure}[tbh]

\subfloat[\label{ 1}]{%
  \includegraphics[width=0.42\linewidth]{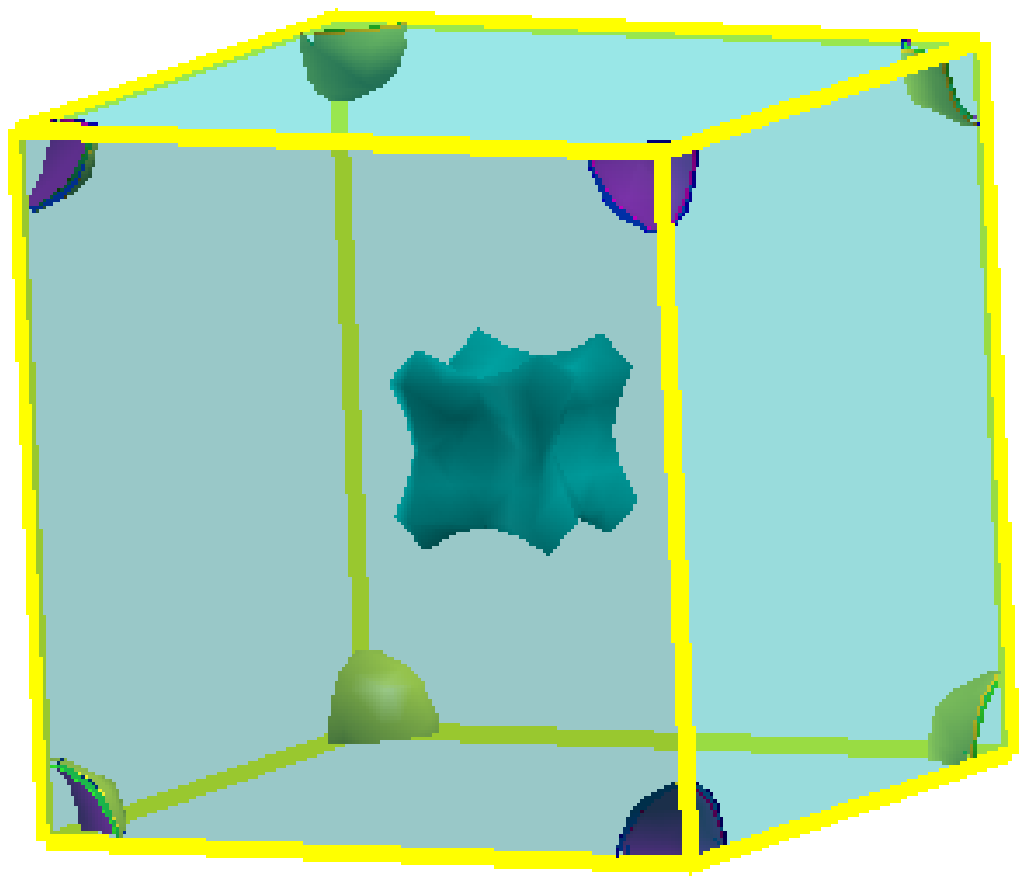}%
}\hfill  
\subfloat[\label{ 2}]{%
  \includegraphics[width=0.42\linewidth]{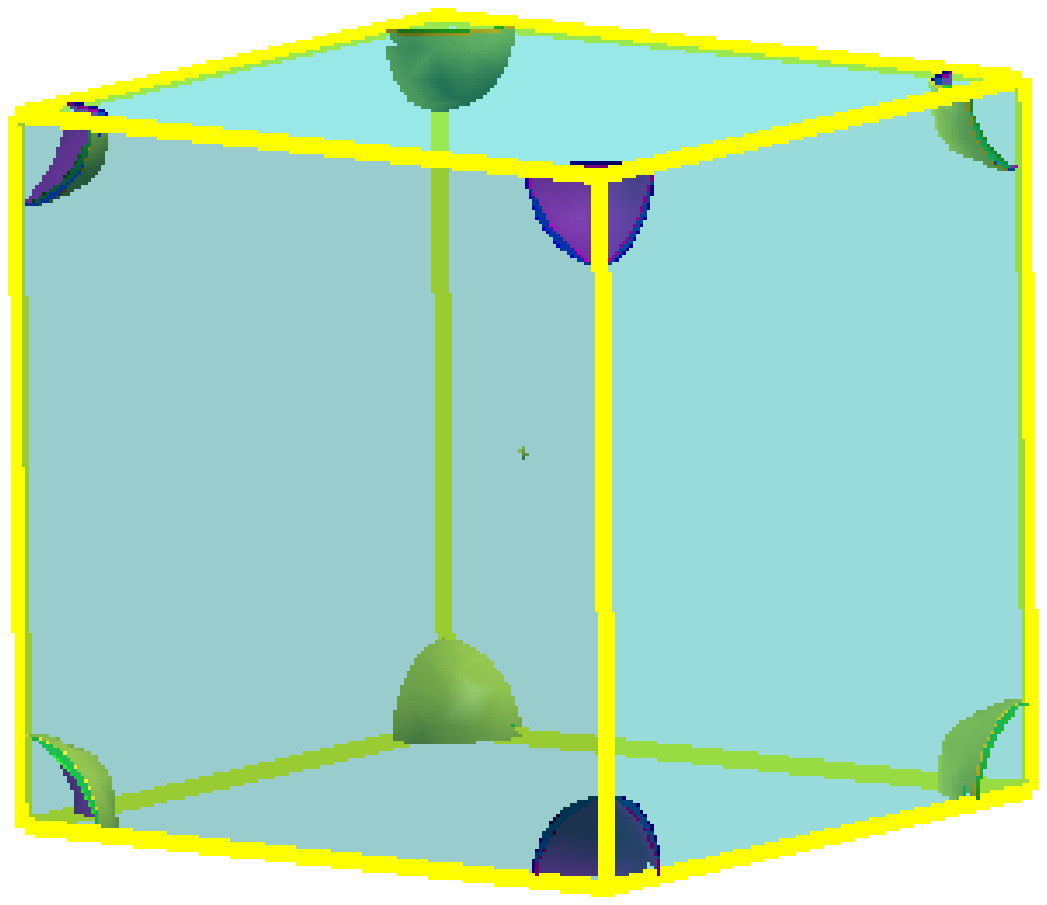}%
}\vfill
\subfloat[\label{ 3}]{%
  \includegraphics[width=0.41\linewidth]{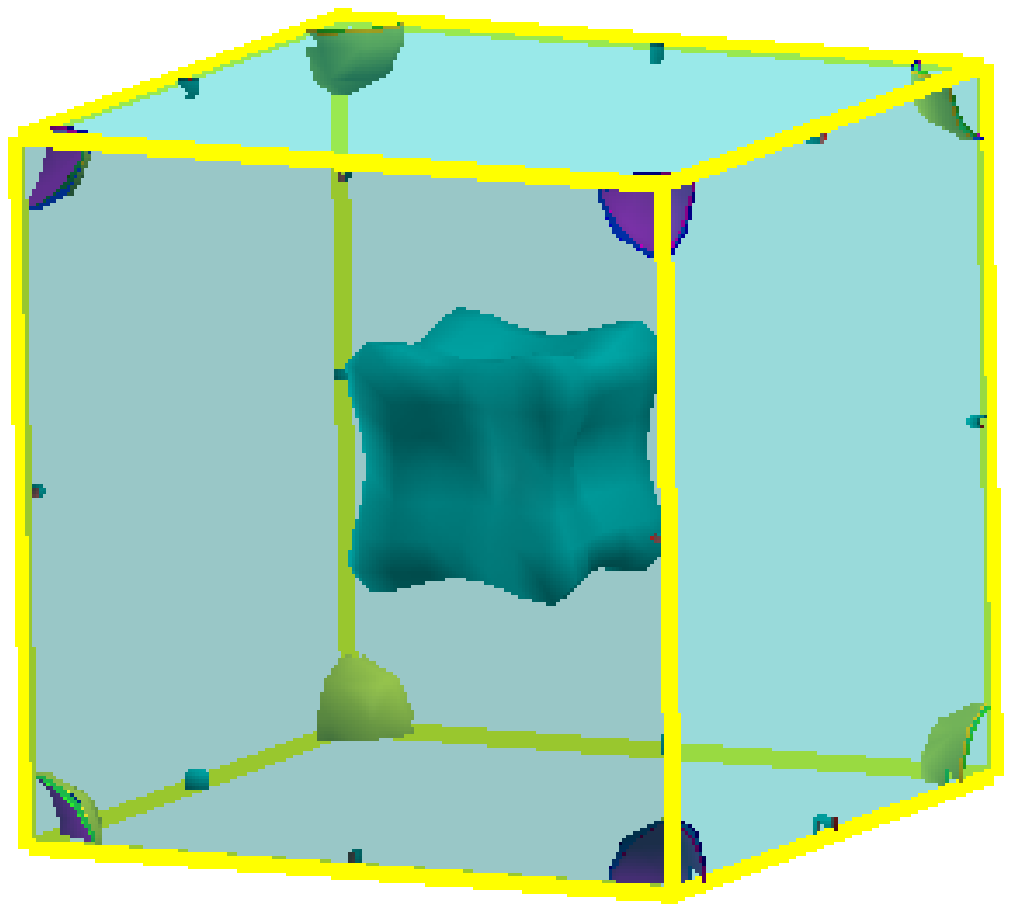}%
}\hfill
\subfloat[\label{ 4}]{%
  \includegraphics[width=0.43\linewidth]{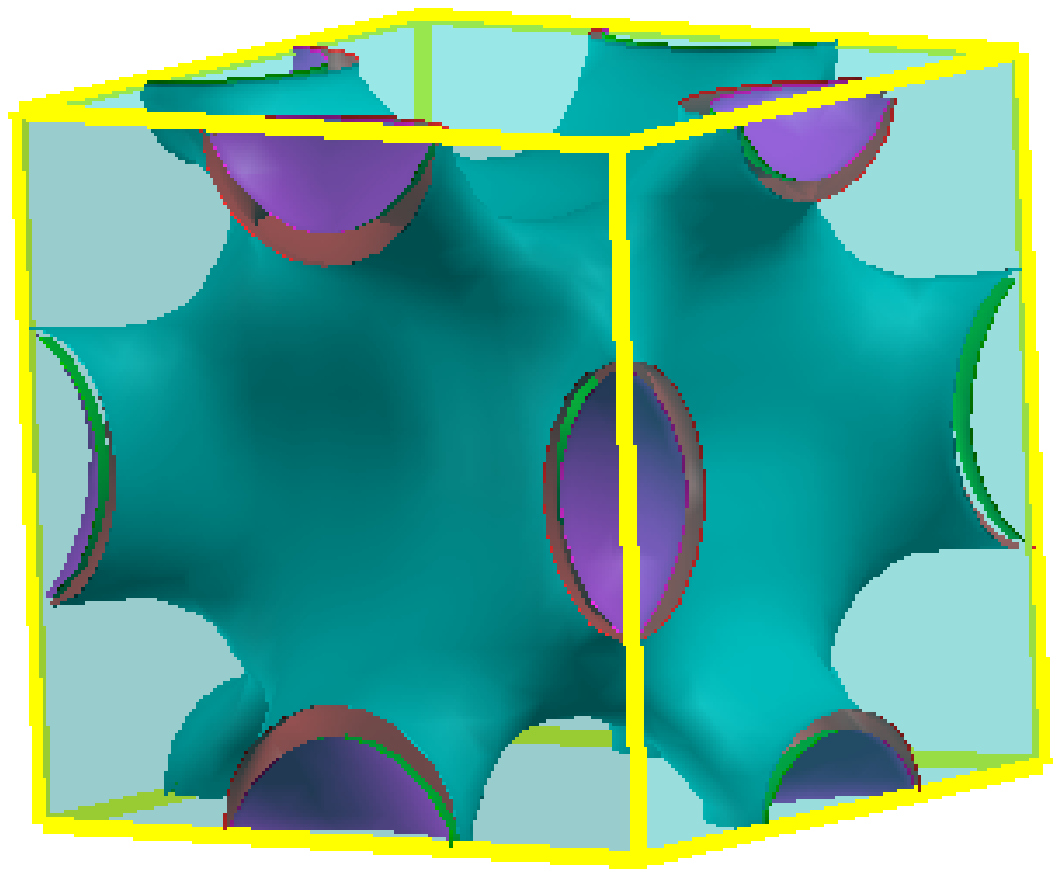}%
}\hfill
\caption{\small{ 3D constant energy surfaces at four different $\mu$ \textit{viz.} (a) $\mu$ at E$_F$, (b) $\mu$ at G1 nodal point, (c) $\mu$ at G2 nodal point, and (d) $\mu$ at R1 nodal point. }}
\label{4}
\end{figure}

The transport coefficients are highly sensitive to the band structure and the available number of states at a given $\mu$. Thus, to understand the behaviour of transport coefficients with T (which will be discussed next) we have also drawn the 3D constant energy surfaces corresponding to E$_F$ and the $\mu$ corresponding to energies of the nodal points (G1, G2 \& R1) which are shown in Fig. 2(a) - 2(d), respectively. These figures correspond to the first Brillouin zone (BZ). It is known that these surfaces are the theoretical area of constant energy in reciprocal space ($k$ space). They define the allowed energies of electrons. These allowed energies of electrons are basically allowed electronic states. Thus, from the figure we tried to get the overview of the allowed electronic states corresponding to the energies of the nodal points and the E$_F$. These figures are directly replicating what we see in Fig. 1. Likewise, in Fig. 2(a) an enclosed structure is seen in the middle of the cube which is $\Gamma$ point, representing the hole pockets while electron pockets can be seen at the vertices of the cube which are $R$ points in the BZ. Furthermore, at the $\mu$ corresponding to the G1 point, the electron pockets are observed only at the vertices. This suggests in the reduction of the electronic states in Fig. 2(b). Further, on moving to $\mu$ corresponding to G2 point, a bigger enclosed structure in the middle of the cube due to the enhanced hole pockets at $\Gamma$ point (as observed in Fig. 1) is found in Fig. 2(c). In addition to this, few small structures can be seen at the centers of the cube edges representing $M$ points in the BZ. This implies increment in the electronic states at the G2 point when compared to E$_F$ and G1 point. Finally, at $\mu$ corresponding to R1 point in Fig. 2(d), the electronic states are found to exist in between these high symmetric points in BZ as observed in Fig. 1. This indicates that the number of electronic states have increased when moving from $\mu$ corresponding to E$_F$ to G2 to R1 points. Though their number get reduced when $\mu$ corresponds to G1 point as compared to other energy surfaces in Fig. 2(a) - 2(d). Thus, from the figure it can be clearly seen that the effective number of states available at $\mu $ corresponding to R1 point is largest whereas it is least at $\mu$ corresponding to G1 point. The effective number of states are seen to follow an increasing trend as R1$>$G2$>$E$_F$ $>$G1. The result and discussion section is further divided into sub-sections where different transport properties at $\mu$ corresponding to the nodal points and the E$_F$ are discussed.

\subsection{Seebeck Coefficient}
\small
\begin{figure}[tbh]
  \begin{center}
   \includegraphics[width=2in]{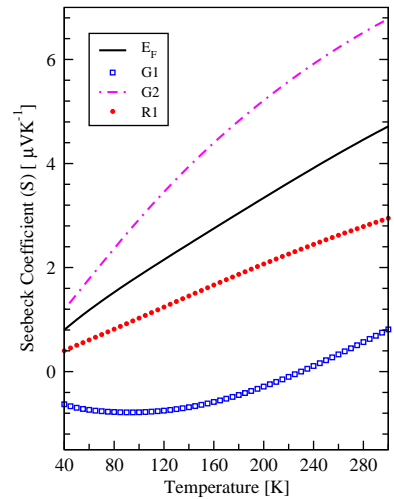}
  \end{center}

  \caption{\small{(colour online) Seebeck coefficient ($S$) versus temperature (T) plot for the T range of 40 - 300 K at four different $\mu$.}}

\end{figure}

Fig. 3 depicts the variation of Seebeck coefficient ($S$) of CoSi within the T range of 40 - 300 K, at four different $\mu$ (E$_F$, G1, G2, \& R1). 
It is seen in the figure that at a given T, the value of \textit{S} at these $\mu$ follows an increasing trend as G2$>$E\textsubscript{F}$>$R1$>$G1. Furthermore, the values of \textit{S} are observed to be positive at the $\mu$ corresponding to E$_F$, G2 \& R1 points. Also, their \textit{S} behaviour is observed to be increasing with rise in T. Their positive \textit{S} values suggests that the dominant charge carriers are holes. However, at the G1 point, it is observed that the value of \textit{S} is initially negative but after a certain T ($>$225 K), it becomes positive. This indicates that for T$<$225 K, the dominant charge carriers are electrons whereas for T$>$225 K, the charge carriers are dominanted by holes. In order to understand this unusual behaviour of \textit{S}, it is important to study the factors upon which it depends. The \textit{S} in solids is defined as\cite{ashcroft} 

%

\begin{equation}
  S=\frac{8\pi^2k_B^2}{3eh^2} T \left(\frac{\pi}{3n}\right)^{2/3}m^*
\end{equation}

where, $k_B$ and $e$ are physical constants representing the Boltzmann constant and the electronic charge respectively. In addition to this the symbol  $m^*$, $n$ and $h$ respectively stands for the effective mass, carrier charge density and the Planck's constant. It is seen from the above formula that the value of \textit{S} gets affected only with the change in the values of $m^*$, T and $n$. The sign of \textit{S} is directly related to $m^*$ because T and $n$ being the positive quantities. Under the parabolic approximation, effective mass is defined as $m^* = \hbar^2/(d^2E/dk^2)$. This implies that the value of $m^*$ is directly related to the curvature of bands. The formula of $m^*$ suggests that its value will be higher for the flat bands in comparision to that of the curved bands. Accordingly, one can also tell about the sign of $m^*$. Likewise, for concave-up bands, its sign will be positive while for concave-down bands, it will be negative. 
Thus, the dispersion plot shown in Fig. 1, seems to play an important role in understanding the behaviour of \textit{S} obtained for the material. It should be noted that in the Fig. 1 many of the bands are almost linear in certain high symmetric directions. There is a possibility that they may deviate from the parabolic behaviour. Therefore, we will try to find the extent upto which the parabolic approximation can help us in understanding the obtained \textit{S} behaviour.

Starting from the G1 point in Fig. 1, there are 4-fold degenerate bands at this point. Among the four bands, two are in CB (almost linear) while the other two are in VB (where one is almost flat and the other is curved). As per the formula of the $m^*$, we cannot say much regarding its value for linear bands. This is because the denominator term in the formula becomes zero for linear bands. However, among the two bands in VB, the flatter band will have higher $m^*$ than the curved ones. Furthermore, it has also been found beforehand that the number of states are least at the $\mu$ corresponding to G1 point. Considering only this factor, it seems that the value of \textit{S} must be largest at the $\mu$ corresponding to G1 point. But the result obtained in Fig. 3 is seen to contradict with our analysis. Therefore, it appears that $m^*$ may play an important role in deciding its transport behaviour for the G1 nodal point. Besides, it is already known that at a finite T electrons get excited thermally from VB, leading to the creation of holes in these bands. Thus, in VB both the charge carriers exist and will contribute to the value of \textit{S}. The close observation of the topmost valence band in the vicinity of G1 point shows that there is a small dip. On moving further down the energy axis, the band becomes flat. Due to the dip in the band, the contribution of the holes to the $m^*$ will be comparatively smaller than that of the electrons. However, as the T increases, the excitations from the flat band will also become effective. As a result, the contributions to the $m^*$ from the holes seems to dominate than that of the electrons. Thus, it is expected that at the lower values of T, the value of \textit{S} will be negative. Beyond a certain T, the values of \textit{S} is then expected to become positive. The observations in Fig. 3 seems to be in terms of the above discussion. The value of T beyond which \textit{S} changes its sign is observed to be 225 K. 
Next, at $\mu$ corresponding to the E$_F$, positive values of \textit{S} is obtained within the given range of T, as seen in Fig. 3. At the $\mu$ corresponding to the Fermi level, it is observed in Fig. 1 that a hole pocket and an electron pocket is present in the vicinity of $\Gamma$ point and $R$ point respectively. Holes from the hole pockets seem to participate more easily than the electrons from electron pocket. This is because of the smaller energy gap corresponding to the hole pockets than that of the electron pocket. This seems to be the reason for getting holes as dominant charge carriers and hence the positive values of \textit{S} here. Then, moving to $\mu$ corresponding to G2 point, it is seen in Fig. 1 that there are three hole pockets, among which two are in $M-\Gamma$ \& $\Gamma-R$ directions and the third one is in the vicinity of $M$ point. In addition to this, there is an electron pocket in the vicinity of $R$ point. At this G2 point, 2-fold degenerate linear bands are observed. Here again, nothing can be said regarding $m^*$ of these linear bands. Yet, one needs to consider factor of $n$ which has higher value at G2 than at the E$_F$ (from Fig. 2). This basically suggests that the value of \textit{S} must be higher at the E$_F$ than at G2 point. But this is contradictory with the result obtained in Fig. 3. This indicates that at this $\mu$ (G2), $m^*$ is dominating over $n$ in the equation (4) which is consequently making its \textit{S} value to be greater than at E$_F$, in Fig. 3. Hence, it appears that linear bands may have large $m^*$ which we are unaware of. This compels that the study of the contribution of linear bands to $m^*$ is needed to be explored. Furthermore, at this $\mu$ there will be both charge carriers participating from VB as discussed before. But, as the values of \textit{S} are positive within the given T range, it seems that the holes are the dominating charge carriers. The possible reason behind this may be the presence of three hole pockets from where the holes can participate more easily than from the electron pocket due to comparatively smaller energy gap. This can also be the reason for the non-linear increment of \textit{S} with the rise in T, for the G2 point in comparison to that of E$_F$. Lastly, for the $\mu$ corresponding to R1 point, 6-fold degenerate bands are observed here (in Fig. 1). It is seen that although these bands are almost linear, they are comparatively more curved than the bands present at G2 point. Due to their linear nature, it becomes difficult to comment over the $m^*$ of these bands. However, it has been already seen that $n$ is largest at the $\mu$ corresponding to R1 point than at any other nodal points (Fig. 2). This suggests that the value of \textit{S} at R1 point must be comparatively smaller than at any other nodal points, which contradicts to the observation of Fig. 3. Moreover, at R1 point, two electron pockets are also present in $R-\Gamma$ and $R-M$ directions as can be seen in Fig. 1. Although, both the charge carriers have the possibility to participate at any finite T. Yet, the observation of positive values of  \textit{S} suggests that the holes are the dominant charge carriers irrespective of the presence of electron pockets. The reason for this behaviour is unclear  and hence more attention is needed. Thus, above discussions show that many of the observations could not be properly explained. This clearly suggests that with the parabolic approximation, the behaviour of \textit{S} for Weyl semi-metals cannot be explained properly. Thus, we do need some other theory which can explain their behaviour appropriately.


\subsection{Electrical Conductivity}


\begin{figure}[tbh]
  \begin{center}
   \includegraphics[width=2.8in]{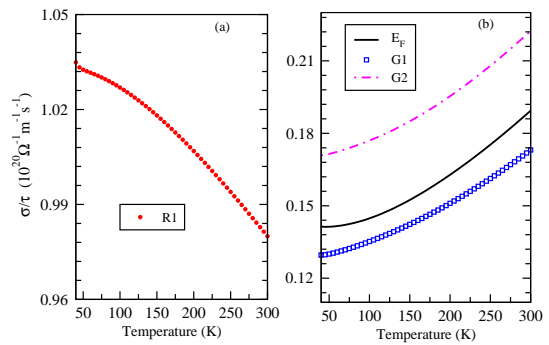}
  \end{center}

  \caption{\small{(colour online) (a) $\boldsymbol \sigma/\tau$ versus temperature (T = 40 - 300 K) plot at R1 nodal point, and (b) at G1, G2 (dashed line) points \& at $\mu$ = zero \textit{i.e.}, Fermi energy (solid line).}}
 
\end{figure}

Now, we discuss the behaviour of electrical conductivity per unit relaxation time ($\boldsymbol \sigma/\tau$) with change in T at different $\mu$, for stoichiometric CoSi. Fig. 4 (a) \& 4 (b) illustrate the plots of the $\boldsymbol \sigma/\tau$ versus T (40 - 300 K) at the four different $\mu$ (E$_F$, G1, G2, \& R1).  From these figures, it is found that \textit{viz.} (i) at any particular T, the value of $\boldsymbol \sigma/\tau$ has shown an increasing behaviour on moving from $\mu$ corresponding to G1 - E$_F$ - G2 - R1. For instance, the value of $\boldsymbol \sigma/\tau$ at T = 100 K is computed to be $\sim$ 0.13$\times$10$^{20}$ $\Omega^{-1} m^{-1} s^{-1}$ when $\mu$ corresponds to G1 whereas its value reaches $ \sim$ 0.18$\times$10$^{20}$ $\Omega^{-1} m^{-1} s^{-1}$ when $\mu$ corresponds to R1. (ii) With the rise in T the value of $\boldsymbol \sigma/\tau$ has shown an increasing behaviour for $\mu$ (E$_F$, G1 \& G2). For explaining these observations, we need to study the expression of the electrical conductivity ($\sigma$) in solids which is given as\cite{ashcroft}, 

\begin{equation}
\boldsymbol \sigma = \boldsymbol \sum_{n} \boldsymbol \sigma_{n} 
\end{equation}  

where,

\begin{equation}
  \boldsymbol{\sigma}_n=e^2 \int_{}^{} \frac{d\boldsymbol k}{4\pi^3} \tau_n(\varepsilon_n(\boldsymbol k)) \boldsymbol v_n(\boldsymbol k)\boldsymbol v_n(\boldsymbol k)\left[-\frac{\partial f}{\partial \varepsilon}\right]_{(\varepsilon_ = \varepsilon_n(\boldsymbol k) ) }
\end{equation}

where, $\epsilon_n(\textbf{\textit{k}})$ is energy, $\tau_n(\textbf{\textit{k}})$ is the relaxation time \& $\boldsymbol v_n(\textbf{\textit{k}})$ is the group velocity of an electron in a level specified by band index $n$ and crystal momentum \textbf{\textit{k}}. Here, $\boldsymbol v_n(\boldsymbol k)$ is defined as \cite{ashcroft},

\begin{equation}
\boldsymbol v_n(\boldsymbol k) = \frac{1}{\hbar} \nabla_{\boldsymbol k} \epsilon_n(\boldsymbol k)
\end{equation}

From the equations (1) to (3), one can see the dependence of $\boldsymbol \sigma$ onto the band structure of a solid. Furthermore, equation (3) shows that $\boldsymbol \sigma_{n}$ depends upon three terms \textit{viz.} (i) $\tau_n(\varepsilon_n(\boldsymbol k)) \boldsymbol v_n(\boldsymbol k)\boldsymbol v_n(\boldsymbol k)$, (ii) $-\frac{\partial f}{\partial \varepsilon}$ at $\epsilon = \epsilon_n(\textbf{\textit{k}})-\mu$, and (iii) number of allowed electronic states available at particular energy. Generally, it is seen that the group velocity ($\boldsymbol v_n(\boldsymbol k)$) does not undergo large variations in the $k_BT$ region. As a result, the velocity is expected to not change to a considerable extent in equation (3). Furthermore, for a given $\mu$, only the energy region having a non-zero value of $-\frac{\partial f}{\partial \varepsilon}$ is considerable. This is because the integrand is significant for only those states which falls in this energy range.  

\begin{figure}[tbh]
  \begin{center}
   \includegraphics[width=2in]{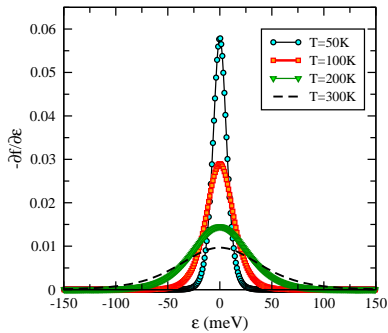}
  \end{center}

  \caption{\small{(colour online) Variation of Fermi-Dirac distribution function with respect to the $\epsilon$($\textbf{\textit{k}}$). }}
 
\end{figure}

Starting from second term, the variations of partial derivative of Fermi-Dirac distribution function ($-\frac{\partial f}{\partial \varepsilon}$) with $\epsilon$ (at $\epsilon = \epsilon_n(\textbf{\textit{k}})-\mu$) are shown in Fig. 5 at four different values of T. These plots are drawn corresponding to $\mu$ equal to 0 meV, which corresponds to E$_F$. Here, it must also be noted that the behaviour of $-\frac{\partial f}{\partial \varepsilon}$ comes out to be similar for other $\mu$. It is seen in the figure that the value of $-\frac{\partial f}{\partial \varepsilon}$ is non-zero only for small energy region around a given $\mu$. This implies that only those $k$-points or the states will contribute to the magnitude of $\boldsymbol \sigma$, which falls in this effective energy region at a given $\mu$. Moving further, it is observed in Fig. 1 that there is no band-gap in the energy range of -0.4 to 0.3 eV around the Fermi level. Thus, sufficient number of states are available in the energy region around all the four $\mu$. Furthermore, it is seen in Fig. 5 that the spread on the energy axis is increasing with the rise in T. Thus, with the increasing T, a greater number of states contribute to the value of $\boldsymbol \sigma$. Hence, it is expected that the value of $\boldsymbol \sigma$ must increase with rise in T. Apart from this, it is already seen in Fig. 2 that the effective number of states available at the different $\mu$ is following an increasing trend as R1 $>$ G2 $>$ E$_F$ $>$G1. Thus, a larger number of states are expected to contribute at the R1 point than at E$_F$ and G2 point at 0 K. This suggests that the value of $\boldsymbol \sigma$ must be highest corresponding to R1 point. On the contrary, least number of states are expected to contribute to the value of $\boldsymbol \sigma$ at the G1 point. This indicates that the value of $\boldsymbol \sigma$ must be least at G1 point for a given T. Accordingly, $\boldsymbol \sigma$ is expected to follow an increasing behaviour as R1 $>$ G2 $>$ E$_F$ $>$G1 at a given T. Thus, the number of available $\textbf{\textit{k}}$ points at a given $\mu$ are playing a major role in deciding the behaviour of $\boldsymbol \sigma$. It is to be mentioned here that the BoltzTraP calculations are carried out under constant relaxation time ($\tau$) approximation. This suggests that the behaviour of $\boldsymbol \sigma/\tau$ must be similar to that of $\boldsymbol \sigma$. Thus, as discussed above, the value of $\boldsymbol \sigma/\tau$ must follow an increasing trend as R1 $>$ G2 $>$ E$_F$ $>$G1 at a given T. In addition to this, at a given $\mu$, the value of $\boldsymbol \sigma/\tau$ must increase with the rise in T. The behaviour obtained in Fig. 4 (b) is analogous to the above discussion. However, we could not understand the reason behind the decreasing behaviour of $\boldsymbol \sigma/\tau$ with the rise in T at $\mu$ corresponding to R1 point in Fig. 4 (a). Thus, it will be quite inetersting to witness these  observations if one performs experiments. 

\subsection{Thermal Conductivity (electronic)}

\begin{figure}[tbh]
  \begin{center}
   \includegraphics[width=1.8in]{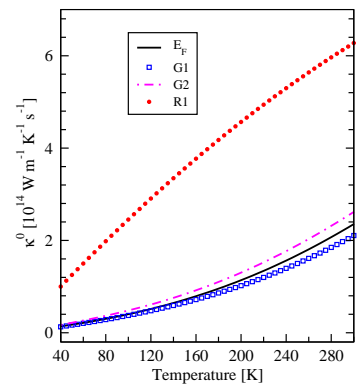}
  \end{center}

  \caption{\small{(colour online) Electronic thermal conductivity per unit relaxation time ($\boldsymbol \kappa_e/\tau=\boldsymbol \kappa^0$) versus temperature (T) plot for the T range of 40 - 300 K at four different $\mu$.}}

\end{figure}

Next, the electronic part of thermal conductivity per unit relaxation time ($\boldsymbol \kappa_e/\tau=\boldsymbol \kappa^0$) is studied for the material corresponding to different $\mu$. Accordingly, the temperature-dependent variation of $\boldsymbol \kappa^0$ with T ranging from 40 - 300 K, at different $\mu$ is shown in Fig. 8. Here again, firstly starting from $\boldsymbol \kappa^0$ behaviour at the $\mu$ corresponding to E$_F$. It is seen to show an increasing behaviour similar to the one reported by Jian \textit{et al.} and Li \textit{et al.} for $\boldsymbol \kappa_e$ \cite{Jian_2020, Li_2005}. From the figure, it is found that, at the E$_F$, the value of $\boldsymbol \kappa^0$ is $\sim$ 2.4 x 10$^{14}$ Wm$^{-1}$K$^{-1}$s$^{-1}$ at 300 K, where 10$^{14}$ is the order of $\tau$. The numerical value \textit{i.e}, 2.4, is seen to be quite close to the experimental result ($\kappa_e\sim$ 3 Wm$^{-1}$K$^{-1}$ at 300 K) \cite{Jian_2020}. Furthermore, the behaviour of $\boldsymbol \kappa^0$ at different nodal points is also observed to be similar to that of the $\mu$ corresponding to E$_F$. At any given T, the maximum value of $\boldsymbol \kappa^0$ has reached when $\mu$ corresponds to R1 point and least when $\mu$ corresponds to G1 point. Thus, the value of $\boldsymbol \kappa^0$ is seen to follow the trend as R1 $>$ G2 $>$ E$_F$ $>$G1. The trend is similar to the one followed by $\boldsymbol \sigma/\tau$ as observed from the Fig. 2. This is because $\boldsymbol \kappa^0$ has a direct relation with $\boldsymbol \sigma/\tau$ which is given by\cite{ashcroft}, 
\begin{equation}
\boldsymbol \kappa^0 = \frac{\pi^2k_B^2}{3e^2} T \big(\frac{\boldsymbol \sigma}{\tau}\big)
\end{equation} 
Here, it is important to note that the temperature-dependent variation of $\boldsymbol \kappa^0$ is obtained to be similar to $\boldsymbol \sigma/\tau$ for $\mu$ corresponding to G1, G2 and E$_F$. This is due to the monotonous increase in the value of $\boldsymbol \sigma/\tau$ with the rise in T at these three points. After seeing their $\boldsymbol \kappa^0$ behaviour, there is possibility to say that $\boldsymbol \sigma/\tau$ term seems to dominate over T term. Although, there is minimal difference in their values of $\boldsymbol \kappa^0$ for any given T. However, the $\boldsymbol \kappa^0$ behaviour at R1 point is quite different from $\boldsymbol \sigma/\tau$ behaviour. $\boldsymbol \kappa^0$ shows an increasing behavior with T having a broad maximum around higher temperature. On seeing this, $\boldsymbol \kappa^0$ value appears to be dominated by T term than $\boldsymbol \sigma/\tau$ in equation (5). As we could not found any experimental evidence of these results (to the best of our knowledge). Thus, there is an urge of performing experiments to visualize these interesting results.


\begin{table}[tbh]
\caption{\footnotesize{Calculated doping concentration required to move the chemical potential to the energies of different nodal points. }}
\label{tab.1}
\begin{center}
\setlength{\tabcolsep}{8pt}
\footnotesize
\begin{tabular}{lccr}
\hline
\hline
 Nodal points & Doping concentration \\
 
  
\hline 
 \\
 G1 (at $\Gamma, \omega\sim$ 30 meV)  & 2.25 $\times$ 10$^{22}$ cm$^{-3}$ (electron doping)\\
 G2 (at $\Gamma, \omega\sim$ -23 meV) & 4.10 $\times$ 10$^{20}$ cm$^{-3}$ (hole doping) \\
 R1 (at $R, \omega \sim$ -186 meV)    & 7.22 $\times$ 10$^{22}$ cm$^{-3}$ (hole doping) \\

\hline
\hline
\end{tabular}
\end{center}
\end{table}

At the end, we have calculated the charge carrier concentration required for doping (electron/hole) to reach the $\mu$ corresponding to the nodal points as shown in the band structure (Fig. 1). The required information has been tabulated in Table I. In the beginning it is said that it will be interesting to study the transport coefficients at nodal points of CoSi. In this regard recent experimental results \cite{Burkov_2019,Antonov_2019,Ovchinnikov_2019} have been found, with which we have tried to relate the theoretical results of this work. In these experimental works, it has been tried to explore the nodal points of CoSi by doping it with the different concentrations of Fe and Ni. In recent work of Dutta \textit{et al.}, they have shown that while moving from Mn-Fe-Co, the bands are found to shift towards the VB due to electron filling of the unoccupied bands \cite{Dutta_2018}. So, there is an expectation that  when it is further moved to Ni, then the nodal points will further shift inside the VB. Therefore, it can be said the Fe doping in CoSi will lower the position of the E$_F$ while the Ni doping in CoSi will tend to higher it. Thus, through this doping it seems that the desired nodal points can be achieved where unconventional electronic topology has been reported. However, it becomes very difficult to achieve a controlled doping technically. Thus, one needs to be extra cautious because these results have already shown that a little change in the concentration can affect the transport behaviour of the material drastically. 

Starting from Antonov \textit{et al.} work where they have reported the experimental results of $\boldsymbol \sigma$ and \textit{S} for Co$_{1-x}$M$_x$Si (M = Fe, Ni; $x$ = 0 - 0.1) \cite{Antonov_2019}. In their work, it is seen that for Co$_{1-x}$Fe$_x$Si ($x \sim$ 0.06 - 0.1) $\boldsymbol \sigma$ is increasing with temperature increment while its rate of change increases with increase in $x$'s value at low temperature region. This increasing trend as followed by $\boldsymbol \sigma$ seems to match with the increasing behaviour of $\boldsymbol \sigma/\tau$ as observed for G2 nodal point in Fig. 4(b). Similarly, for Co$_{1-x}$Ni$_x$Si alloy $\boldsymbol \sigma$ increases till 400 K with a very small rate which seems to be similar to increasing behaviour of $\boldsymbol \sigma/\tau$ for G1 nodal point (as observed in Fig. 4(b)) when  $ x \sim$ 0.01 \cite{Antonov_2019}. Further in their work, the positive \textit{S} (increasing behaviour) for Co$_{1-x}$Fe$_x$Si ($x \sim$ 0.06 - 0.1) seems to go with our \textit{S} result for G2 and R1 points in Fig. 3. When $x$ varies from 0 to 0.03 in Co$_{1-x}$Ni$_x$Si, there is a possibility of witnessing the change in the charge dominance from electrons to holes \cite{Antonov_2019} similar to our G1's \textit{S} result. Thus, based on these results \cite{Antonov_2019}, it seems for achieving R1's result, $x$ needs to be $0.1$ (or more) and for G2 (x $\sim$ 0.06 - 0.1) when doped with Fe$_x$, and for G1 ($ x \sim$ 0.01) when doped with Ni$_x$. Similar inference can be made from Burkov \textit{et al.} results \cite{Burkov_2019}. So, next moving to Ovchinnikov \textit{et al.} work, they have studied the electrical $\boldsymbol \sigma$ of Co$_{1-x}$M$_x$Si with Fe and Ni where $x$ ranges upto 8 and 5 at. \%, respectively\cite{Ovchinnikov_2019}. Here again, their increasing $\boldsymbol \sigma$ behaviour for Co$_{1-x}$Fe$_{x}$Si ($x$ = 0.015, 0.04 \& 0.05) seems similar to $\boldsymbol \sigma/\tau$ behaviour of G2 point as observed in Fig. 4(b). Yet, their $\boldsymbol \sigma$ behaviour for Co$_{0.95}$Ni$_{0.05}$Si does not match with G1 point's $\boldsymbol \sigma/\tau$ result in Fig. 4(b). Thus, on seeing both experimental works, it seems there is a discrepancy when compared with each other results but they are found to be fairly in agreement with our theoretical results. This mismatching of their results might be due to the difficulty arises in controlled doping which is most important in studying those nodal points' transport properties. However, there are still some points which could not be matched and discussed in this study with experimental results. Here it is important to note that we have not considered $\tau$ for the discussion through out the study and that might be the possible reason. This makes the urge for low-temperature transport coefficients of CoSi-based alloys for further investigation experimentally.       

\section{Conclusion}

CoSi has been predicted to have three types of linear unconventional band crossings, two at $\Gamma$ (G1 \& G2) and one at $R$ (R1) points in Brillouin zone. At these band-crossing/nodal points, new fermions have been reported to be present. 
In this work we have presented a study where various low-temperature dependent transport coefficients (\textit{S}, $\boldsymbol \sigma/\tau$ and $\boldsymbol \kappa^0$) are studied using DFT method within the temperature (T) range of 40 - 300 K. For understanding their behaviours, the constant energy surfaces are constructed corresponding to energies of these nodal points and the E$_F$. They have shown that the number of states available at their respective energies follow an increasing behavior as R1 $>$ G2 $>$ E$_F$ $>$ G1 at T = 0 K.  Similar behaviour appears to be observed by other transport coefficients at these nodal points and at E$_F$ with the rise in T such as $\boldsymbol \sigma/\tau$ and $\boldsymbol \kappa^0$. However, \textit{S} appears to follow an increasing trend as G2 $>$ E$_F$ $>$ R1 $>$ G1. It has a positive values at G2, R1 and E$_F$ within the given range of T. But at the G1 point, it has negative values for T $<$ 225 K and beyond 225 K, it changes to positive value. The low temperature-dependence of $\boldsymbol \sigma/\tau$ and \textit{S} at these nodal points seem to follow the behaviour observed in recent experimental works (not magnitude wise). It is varying from one experimental work to the other. This suggests that an appropriate concentration is very important to capture these unconventional electronic transport. In this regard, we have calculated the required doping concentrations at 300 K such as for G1 point (electron doping $\sim$ 2.25 $\times$ 10$^{22}$ cm$^{-3}$), G2 \& R1 points (hole doping $\sim$ 4.10 $\times$ 10$^{20}$ cm$^{-3}$ \& 7.22 $\times$ 10$^{22}$ cm$^{-3}$, respectively). 

\bibliography{MS}

\begin{thebibliography}{47}%
\makeatletter
\providecommand \@ifxundefined [1]{%
 \@ifx{#1\undefined}
}%
\providecommand \@ifnum [1]{%
 \ifnum #1\expandafter \@firstoftwo
 \else \expandafter \@secondoftwo
 \fi
}%
\providecommand \@ifx [1]{%
 \ifx #1\expandafter \@firstoftwo
 \else \expandafter \@secondoftwo
 \fi
}%
\providecommand \natexlab [1]{#1}%
\providecommand \enquote  [1]{``#1''}%
\providecommand \bibnamefont  [1]{#1}%
\providecommand \bibfnamefont [1]{#1}%
\providecommand \citenamefont [1]{#1}%
\providecommand \href@noop [0]{\@secondoftwo}%
\providecommand \href [0]{\begingroup \@sanitize@url \@href}%
\providecommand \@href[1]{\@@startlink{#1}\@@href}%
\providecommand \@@href[1]{\endgroup#1\@@endlink}%
\providecommand \@sanitize@url [0]{\catcode `\\12\catcode `\$12\catcode
  `\&12\catcode `\#12\catcode `\^12\catcode `\_12\catcode `\%12\relax}%
\providecommand \@@startlink[1]{}%
\providecommand \@@endlink[0]{}%
\providecommand \url  [0]{\begingroup\@sanitize@url \@url }%
\providecommand \@url [1]{\endgroup\@href {#1}{\urlprefix }}%
\providecommand \urlprefix  [0]{URL }%
\providecommand \Eprint [0]{\href }%
\providecommand \doibase [0]{http://dx.doi.org/}%
\providecommand \selectlanguage [0]{\@gobble}%
\providecommand \bibinfo  [0]{\@secondoftwo}%
\providecommand \bibfield  [0]{\@secondoftwo}%
\providecommand \translation [1]{[#1]}%
\providecommand \BibitemOpen [0]{}%
\providecommand \bibitemStop [0]{}%
\providecommand \bibitemNoStop [0]{.\EOS\space}%
\providecommand \EOS [0]{\spacefactor3000\relax}%
\providecommand \BibitemShut  [1]{\csname bibitem#1\endcsname}%
\let\auto@bib@innerbib\@empty
\bibitem [{\citenamefont {Lange}(1997)}]{Lange_1997}%
  \BibitemOpen
  \bibfield  {author} {\bibinfo {author} {\bibfnamefont {H.}~\bibnamefont
  {Lange}},\ }\href@noop {} {\bibfield  {journal} {\bibinfo  {journal} {physica
  status solidi (b)}\ }\textbf {\bibinfo {volume} {201}},\ \bibinfo {pages} {3}
  (\bibinfo {year} {1997})}\BibitemShut {NoStop}%
\bibitem [{\citenamefont {Fedrov}\ and\ \citenamefont
  {Zaitsev}(1995)}]{Fedrov_1995}%
  \BibitemOpen
  \bibfield  {author} {\bibinfo {author} {\bibfnamefont {M.~I.}\ \bibnamefont
  {Fedrov}}\ and\ \bibinfo {author} {\bibfnamefont {V.~K.}\ \bibnamefont
  {Zaitsev}},\ }\href@noop {} {\emph {\bibinfo {title} {CRC Handbook of
  Thermoelectrics}}},\ edited by\ \bibinfo {editor} {\bibfnamefont {D.~M.}\
  \bibnamefont {Rowe}}\ (\bibinfo  {publisher} {Boca Raton, FL : CRC Press},\
  \bibinfo {year} {1995})\ Chap.~\bibinfo {chapter} {27}, pp.\ \bibinfo {pages}
  {321--328}\BibitemShut {NoStop}%
\bibitem [{\citenamefont {Asanabe}\ \emph {et~al.}(1964)\citenamefont
  {Asanabe}, \citenamefont {Shinoda},\ and\ \citenamefont
  {Sasaki}}]{Asanabe_1964}%
  \BibitemOpen
  \bibfield  {author} {\bibinfo {author} {\bibfnamefont {S.}~\bibnamefont
  {Asanabe}}, \bibinfo {author} {\bibfnamefont {D.}~\bibnamefont {Shinoda}}, \
  and\ \bibinfo {author} {\bibfnamefont {Y.}~\bibnamefont {Sasaki}},\
  }\href@noop {} {\bibfield  {journal} {\bibinfo  {journal} {Phys. Rev.}\
  }\textbf {\bibinfo {volume} {134}},\ \bibinfo {pages} {A774} (\bibinfo {year}
  {1964})}\BibitemShut {NoStop}%
\bibitem [{\citenamefont {Kim}\ \emph {et~al.}(2002)\citenamefont {Kim},
  \citenamefont {Mishima},\ and\ \citenamefont {Choi}}]{Kim_2002}%
  \BibitemOpen
  \bibfield  {author} {\bibinfo {author} {\bibfnamefont {S.}~\bibnamefont
  {Kim}}, \bibinfo {author} {\bibfnamefont {Y.}~\bibnamefont {Mishima}}, \ and\
  \bibinfo {author} {\bibfnamefont {D.}~\bibnamefont {Choi}},\ }\href@noop {}
  {\bibfield  {journal} {\bibinfo  {journal} {Intermetallics}\ }\textbf
  {\bibinfo {volume} {10}},\ \bibinfo {pages} {177} (\bibinfo {year}
  {2002})}\BibitemShut {NoStop}%
\bibitem [{\citenamefont {Nikitin}\ \emph {et~al.}(1970)\citenamefont
  {Nikitin}, \citenamefont {Tamarin},\ and\ \citenamefont
  {Tarasov}}]{Nikitin_1970}%
  \BibitemOpen
  \bibfield  {author} {\bibinfo {author} {\bibfnamefont {E.~N.}\ \bibnamefont
  {Nikitin}}, \bibinfo {author} {\bibfnamefont {P.~V.}\ \bibnamefont
  {Tamarin}}, \ and\ \bibinfo {author} {\bibfnamefont {V.~I.}\ \bibnamefont
  {Tarasov}},\ }\href@noop {} {\bibfield  {journal} {\bibinfo  {journal} {Sov.
  Phys. Solid State}\ }\textbf {\bibinfo {volume} {11}},\ \bibinfo {pages}
  {2002} (\bibinfo {year} {1970})}\BibitemShut {NoStop}%
\bibitem [{\citenamefont {Alekseeva}\ \emph {et~al.}(1981)\citenamefont
  {Alekseeva}, \citenamefont {Zailsev}, \citenamefont {Petrov}, \citenamefont
  {Tarasov},\ and\ \citenamefont {Fedorov}}]{Alekseeva_1981}%
  \BibitemOpen
  \bibfield  {author} {\bibinfo {author} {\bibfnamefont {G.~T.}\ \bibnamefont
  {Alekseeva}}, \bibinfo {author} {\bibfnamefont {V.~K.}\ \bibnamefont
  {Zailsev}}, \bibinfo {author} {\bibfnamefont {A.~V.}\ \bibnamefont {Petrov}},
  \bibinfo {author} {\bibfnamefont {V.~I.}\ \bibnamefont {Tarasov}}, \ and\
  \bibinfo {author} {\bibfnamefont {M.~I.}\ \bibnamefont {Fedorov}},\
  }\href@noop {} {\bibfield  {journal} {\bibinfo  {journal} {Sov. Phys. Solid
  State}\ }\textbf {\bibinfo {volume} {23}},\ \bibinfo {pages} {1685} (\bibinfo
  {year} {1981})}\BibitemShut {NoStop}%
\bibitem [{\citenamefont {Liu}\ \emph {et~al.}(2012)\citenamefont {Liu},
  \citenamefont {Li},\ and\ \citenamefont {Wang}}]{Liu_2012}%
  \BibitemOpen
  \bibfield  {author} {\bibinfo {author} {\bibfnamefont {Y.}~\bibnamefont
  {Liu}}, \bibinfo {author} {\bibfnamefont {S.-N.}\ \bibnamefont {Li}}, \ and\
  \bibinfo {author} {\bibfnamefont {Z.-Z.}\ \bibnamefont {Wang}},\ }\href@noop
  {} {\bibfield  {journal} {\bibinfo  {journal} {Physica B: Condensed Matter}\
  }\textbf {\bibinfo {volume} {407}},\ \bibinfo {pages} {4700} (\bibinfo {year}
  {2012})}\BibitemShut {NoStop}%
\bibitem [{\citenamefont {Lue}\ \emph {et~al.}(2004)\citenamefont {Lue},
  \citenamefont {Kuo}, \citenamefont {Huang},\ and\ \citenamefont
  {Lai}}]{Lue_2004}%
  \BibitemOpen
  \bibfield  {author} {\bibinfo {author} {\bibfnamefont {C.~S.}\ \bibnamefont
  {Lue}}, \bibinfo {author} {\bibfnamefont {Y.-K.}\ \bibnamefont {Kuo}},
  \bibinfo {author} {\bibfnamefont {C.~L.}\ \bibnamefont {Huang}}, \ and\
  \bibinfo {author} {\bibfnamefont {W.~J.}\ \bibnamefont {Lai}},\ }\href@noop
  {} {\bibfield  {journal} {\bibinfo  {journal} {Phys. Rev. B}\ }\textbf
  {\bibinfo {volume} {69}},\ \bibinfo {pages} {125111} (\bibinfo {year}
  {2004})}\BibitemShut {NoStop}%
\bibitem [{\citenamefont {Sakai}\ \emph {et~al.}(2007)\citenamefont {Sakai},
  \citenamefont {Ishii}, \citenamefont {Onose}, \citenamefont {Tomioka},
  \citenamefont {Yotsuhashi}, \citenamefont {Adachi}, \citenamefont {Nagaosa},\
  and\ \citenamefont {Tokura}}]{Sakai_2007}%
  \BibitemOpen
  \bibfield  {author} {\bibinfo {author} {\bibfnamefont {A.}~\bibnamefont
  {Sakai}}, \bibinfo {author} {\bibfnamefont {F.}~\bibnamefont {Ishii}},
  \bibinfo {author} {\bibfnamefont {Y.}~\bibnamefont {Onose}}, \bibinfo
  {author} {\bibfnamefont {Y.}~\bibnamefont {Tomioka}}, \bibinfo {author}
  {\bibfnamefont {S.}~\bibnamefont {Yotsuhashi}}, \bibinfo {author}
  {\bibfnamefont {H.}~\bibnamefont {Adachi}}, \bibinfo {author} {\bibfnamefont
  {N.}~\bibnamefont {Nagaosa}}, \ and\ \bibinfo {author} {\bibfnamefont
  {Y.}~\bibnamefont {Tokura}},\ }\href@noop {} {\bibfield  {journal} {\bibinfo
  {journal} {J. Phys. Soc. Jpn.}\ }\textbf {\bibinfo {volume} {76}},\ \bibinfo
  {pages} {093601} (\bibinfo {year} {2007})}\BibitemShut {NoStop}%
\bibitem [{\citenamefont {Rowe}(1995)}]{Rowe_1995}%
  \BibitemOpen
  \bibfield  {author} {\bibinfo {author} {\bibfnamefont {D.~M.}\ \bibnamefont
  {Rowe}},\ }\href@noop {} {\emph {\bibinfo {title} {CRC Handbook of
  Thermoelectrics}}}\ (\bibinfo  {publisher} {Boca Raton, FL : CRC Press},\
  \bibinfo {year} {1995})\BibitemShut {NoStop}%
\bibitem [{\citenamefont {Imai}\ \emph {et~al.}(2001)\citenamefont {Imai},
  \citenamefont {Mukaida}, \citenamefont {Kobayashi},\ and\ \citenamefont
  {Tsunoda}}]{Imai_2001}%
  \BibitemOpen
  \bibfield  {author} {\bibinfo {author} {\bibfnamefont {Y.}~\bibnamefont
  {Imai}}, \bibinfo {author} {\bibfnamefont {M.}~\bibnamefont {Mukaida}},
  \bibinfo {author} {\bibfnamefont {K.}~\bibnamefont {Kobayashi}}, \ and\
  \bibinfo {author} {\bibfnamefont {T.}~\bibnamefont {Tsunoda}},\ }\href@noop
  {} {\bibfield  {journal} {\bibinfo  {journal} {Intermetallics}\ }\textbf
  {\bibinfo {volume} {9}},\ \bibinfo {pages} {261} (\bibinfo {year}
  {2001})}\BibitemShut {NoStop}%
\bibitem [{\citenamefont {Pan}\ \emph {et~al.}(2007)\citenamefont {Pan},
  \citenamefont {Zhang},\ and\ \citenamefont {Wu}}]{Pan_2007}%
  \BibitemOpen
  \bibfield  {author} {\bibinfo {author} {\bibfnamefont {Z.~J.}\ \bibnamefont
  {Pan}}, \bibinfo {author} {\bibfnamefont {L.~T.}\ \bibnamefont {Zhang}}, \
  and\ \bibinfo {author} {\bibfnamefont {J.~S.}\ \bibnamefont {Wu}},\
  }\href@noop {} {\bibfield  {journal} {\bibinfo  {journal} {Journal of Applied
  Physics}\ }\textbf {\bibinfo {volume} {101}},\ \bibinfo {pages} {033715}
  (\bibinfo {year} {2007})}\BibitemShut {NoStop}%
\bibitem [{\citenamefont {{Benoit, R.}}(1955)}]{Benoit_1955}%
  \BibitemOpen
  \bibfield  {author} {\bibinfo {author} {\bibnamefont {{Benoit, R.}}},\
  }\href@noop {} {\bibfield  {journal} {\bibinfo  {journal} {J. Chim. Phys.}\
  }\textbf {\bibinfo {volume} {52}},\ \bibinfo {pages} {119} (\bibinfo {year}
  {1955})}\BibitemShut {NoStop}%
\bibitem [{\citenamefont {Wertheim}\ \emph {et~al.}(1966)\citenamefont
  {Wertheim}, \citenamefont {Wernick},\ and\ \citenamefont
  {Buchanan}}]{Wertheim_1966}%
  \BibitemOpen
  \bibfield  {author} {\bibinfo {author} {\bibfnamefont {G.~K.}\ \bibnamefont
  {Wertheim}}, \bibinfo {author} {\bibfnamefont {J.~H.}\ \bibnamefont
  {Wernick}}, \ and\ \bibinfo {author} {\bibfnamefont {D.~N.~E.}\ \bibnamefont
  {Buchanan}},\ }\href@noop {} {\bibfield  {journal} {\bibinfo  {journal}
  {Journal of Applied Physics}\ }\textbf {\bibinfo {volume} {37}},\ \bibinfo
  {pages} {3333} (\bibinfo {year} {1966})}\BibitemShut {NoStop}%
\bibitem [{\citenamefont {Wernick}\ \emph {et~al.}(1972)\citenamefont
  {Wernick}, \citenamefont {Wertheim},\ and\ \citenamefont
  {Sherwood}}]{Wertheim_1972}%
  \BibitemOpen
  \bibfield  {author} {\bibinfo {author} {\bibfnamefont {J.}~\bibnamefont
  {Wernick}}, \bibinfo {author} {\bibfnamefont {G.}~\bibnamefont {Wertheim}}, \
  and\ \bibinfo {author} {\bibfnamefont {R.}~\bibnamefont {Sherwood}},\
  }\href@noop {} {\bibfield  {journal} {\bibinfo  {journal} {Materials Research
  Bulletin}\ }\textbf {\bibinfo {volume} {7}},\ \bibinfo {pages} {1431}
  (\bibinfo {year} {1972})}\BibitemShut {NoStop}%
\bibitem [{\citenamefont {Petrova}\ \emph {et~al.}(2010)\citenamefont
  {Petrova}, \citenamefont {Krasnorussky}, \citenamefont {Shikov},
  \citenamefont {Yuhasz}, \citenamefont {Lograsso}, \citenamefont {Lashley},\
  and\ \citenamefont {Stishov}}]{Petrova_2010}%
  \BibitemOpen
  \bibfield  {author} {\bibinfo {author} {\bibfnamefont {A.~E.}\ \bibnamefont
  {Petrova}}, \bibinfo {author} {\bibfnamefont {V.~N.}\ \bibnamefont
  {Krasnorussky}}, \bibinfo {author} {\bibfnamefont {A.~A.}\ \bibnamefont
  {Shikov}}, \bibinfo {author} {\bibfnamefont {W.~M.}\ \bibnamefont {Yuhasz}},
  \bibinfo {author} {\bibfnamefont {T.~A.}\ \bibnamefont {Lograsso}}, \bibinfo
  {author} {\bibfnamefont {J.~C.}\ \bibnamefont {Lashley}}, \ and\ \bibinfo
  {author} {\bibfnamefont {S.~M.}\ \bibnamefont {Stishov}},\ }\href@noop {}
  {\bibfield  {journal} {\bibinfo  {journal} {Phys. Rev. B}\ }\textbf {\bibinfo
  {volume} {82}},\ \bibinfo {pages} {155124} (\bibinfo {year}
  {2010})}\BibitemShut {NoStop}%
\bibitem [{\citenamefont {Amamou}\ \emph {et~al.}(1972)\citenamefont {Amamou},
  \citenamefont {Bach}, \citenamefont {Gautier}, \citenamefont {Robert},\ and\
  \citenamefont {Castaing}}]{Amamou_1972}%
  \BibitemOpen
  \bibfield  {author} {\bibinfo {author} {\bibfnamefont {A.}~\bibnamefont
  {Amamou}}, \bibinfo {author} {\bibfnamefont {P.}~\bibnamefont {Bach}},
  \bibinfo {author} {\bibfnamefont {F.}~\bibnamefont {Gautier}}, \bibinfo
  {author} {\bibfnamefont {C.}~\bibnamefont {Robert}}, \ and\ \bibinfo {author}
  {\bibfnamefont {J.}~\bibnamefont {Castaing}},\ }\href@noop {} {\bibfield
  {journal} {\bibinfo  {journal} {Journal of Physics and Chemistry of Solids}\
  }\textbf {\bibinfo {volume} {33}},\ \bibinfo {pages} {1697} (\bibinfo {year}
  {1972})}\BibitemShut {NoStop}%
\bibitem [{\citenamefont {Ishii}\ \emph {et~al.}(2014)\citenamefont {Ishii},
  \citenamefont {Kotaka},\ and\ \citenamefont {Onishi}}]{Ishii_2014}%
  \BibitemOpen
  \bibfield  {author} {\bibinfo {author} {\bibfnamefont {F.}~\bibnamefont
  {Ishii}}, \bibinfo {author} {\bibfnamefont {H.}~\bibnamefont {Kotaka}}, \
  and\ \bibinfo {author} {\bibfnamefont {T.}~\bibnamefont {Onishi}},\
  }\href@noop {} {\bibfield  {journal} {\bibinfo  {journal} {JPS Conf. Proc.}\
  }\textbf {\bibinfo {volume} {3}},\ \bibinfo {pages} {016019} (\bibinfo {year}
  {2014})}\BibitemShut {NoStop}%
\bibitem [{\citenamefont {Wehling}\ \emph {et~al.}(2014)\citenamefont
  {Wehling}, \citenamefont {Black-Schaffer},\ and\ \citenamefont
  {Balatsky}}]{Wehling_2014}%
  \BibitemOpen
  \bibfield  {author} {\bibinfo {author} {\bibfnamefont {T.}~\bibnamefont
  {Wehling}}, \bibinfo {author} {\bibfnamefont {A.}~\bibnamefont
  {Black-Schaffer}}, \ and\ \bibinfo {author} {\bibfnamefont {A.}~\bibnamefont
  {Balatsky}},\ }\href@noop {} {\bibfield  {journal} {\bibinfo  {journal}
  {Advances in Physics}\ }\textbf {\bibinfo {volume} {63}},\ \bibinfo {pages}
  {1} (\bibinfo {year} {2014})}\BibitemShut {NoStop}%
\bibitem [{\citenamefont {Huang}\ \emph {et~al.}(2016)\citenamefont {Huang},
  \citenamefont {Xu}, \citenamefont {Belopolski}, \citenamefont {Lee},
  \citenamefont {Chang}, \citenamefont {Chang}, \citenamefont {Wang},
  \citenamefont {Alidoust}, \citenamefont {Bian}, \citenamefont {Neupane},
  \citenamefont {Sanchez}, \citenamefont {Zheng}, \citenamefont {Jeng},
  \citenamefont {Bansil}, \citenamefont {Neupert}, \citenamefont {Lin},\ and\
  \citenamefont {Hasan}}]{Huang_2016}%
  \BibitemOpen
  \bibfield  {author} {\bibinfo {author} {\bibfnamefont {S.-M.}\ \bibnamefont
  {Huang}}, \bibinfo {author} {\bibfnamefont {S.-Y.}\ \bibnamefont {Xu}},
  \bibinfo {author} {\bibfnamefont {I.}~\bibnamefont {Belopolski}}, \bibinfo
  {author} {\bibfnamefont {C.-C.}\ \bibnamefont {Lee}}, \bibinfo {author}
  {\bibfnamefont {G.}~\bibnamefont {Chang}}, \bibinfo {author} {\bibfnamefont
  {T.-R.}\ \bibnamefont {Chang}}, \bibinfo {author} {\bibfnamefont
  {B.}~\bibnamefont {Wang}}, \bibinfo {author} {\bibfnamefont {N.}~\bibnamefont
  {Alidoust}}, \bibinfo {author} {\bibfnamefont {G.}~\bibnamefont {Bian}},
  \bibinfo {author} {\bibfnamefont {M.}~\bibnamefont {Neupane}}, \bibinfo
  {author} {\bibfnamefont {D.}~\bibnamefont {Sanchez}}, \bibinfo {author}
  {\bibfnamefont {H.}~\bibnamefont {Zheng}}, \bibinfo {author} {\bibfnamefont
  {H.-T.}\ \bibnamefont {Jeng}}, \bibinfo {author} {\bibfnamefont
  {A.}~\bibnamefont {Bansil}}, \bibinfo {author} {\bibfnamefont
  {T.}~\bibnamefont {Neupert}}, \bibinfo {author} {\bibfnamefont
  {H.}~\bibnamefont {Lin}}, \ and\ \bibinfo {author} {\bibfnamefont {M.~Z.}\
  \bibnamefont {Hasan}},\ }\href {\doibase 10.1073/pnas.1514581113} {\bibfield
  {journal} {\bibinfo  {journal} {Proc. Natl. Acad. Sci.}\ }\textbf {\bibinfo
  {volume} {113}},\ \bibinfo {pages} {1180} (\bibinfo {year}
  {2016})}\BibitemShut {NoStop}%
\bibitem [{\citenamefont {Pauling}\ and\ \citenamefont
  {Soldate}(1948)}]{Pauling_1948}%
  \BibitemOpen
  \bibfield  {author} {\bibinfo {author} {\bibfnamefont {L.}~\bibnamefont
  {Pauling}}\ and\ \bibinfo {author} {\bibfnamefont {A.~M.}\ \bibnamefont
  {Soldate}},\ }\href@noop {} {\bibfield  {journal} {\bibinfo  {journal} {Acta
  Crystallographica}\ }\textbf {\bibinfo {volume} {1}},\ \bibinfo {pages} {212}
  (\bibinfo {year} {1948})}\BibitemShut {NoStop}%
\bibitem [{\citenamefont {Bradlyn}\ \emph {et~al.}(2016)\citenamefont
  {Bradlyn}, \citenamefont {Cano}, \citenamefont {Wang}, \citenamefont
  {Vergniory}, \citenamefont {Felser}, \citenamefont {Cava},\ and\
  \citenamefont {Bernevig}}]{Bradlyn_2016}%
  \BibitemOpen
  \bibfield  {author} {\bibinfo {author} {\bibfnamefont {B.}~\bibnamefont
  {Bradlyn}}, \bibinfo {author} {\bibfnamefont {J.}~\bibnamefont {Cano}},
  \bibinfo {author} {\bibfnamefont {Z.}~\bibnamefont {Wang}}, \bibinfo {author}
  {\bibfnamefont {M.~G.}\ \bibnamefont {Vergniory}}, \bibinfo {author}
  {\bibfnamefont {C.}~\bibnamefont {Felser}}, \bibinfo {author} {\bibfnamefont
  {R.~J.}\ \bibnamefont {Cava}}, \ and\ \bibinfo {author} {\bibfnamefont
  {B.~A.}\ \bibnamefont {Bernevig}},\ }\href@noop {} {\bibfield  {journal}
  {\bibinfo  {journal} {Science}\ }\textbf {\bibinfo {volume} {353}} (\bibinfo
  {year} {2016})}\BibitemShut {NoStop}%
\bibitem [{\citenamefont {Tang}\ \emph {et~al.}(2017)\citenamefont {Tang},
  \citenamefont {Zhou},\ and\ \citenamefont {Zhang}}]{Tang_2017}%
  \BibitemOpen
  \bibfield  {author} {\bibinfo {author} {\bibfnamefont {P.}~\bibnamefont
  {Tang}}, \bibinfo {author} {\bibfnamefont {Q.}~\bibnamefont {Zhou}}, \ and\
  \bibinfo {author} {\bibfnamefont {S.-C.}\ \bibnamefont {Zhang}},\ }\href@noop
  {} {\bibfield  {journal} {\bibinfo  {journal} {Phys. Rev. Lett.}\ }\textbf
  {\bibinfo {volume} {119}},\ \bibinfo {pages} {206402} (\bibinfo {year}
  {2017})}\BibitemShut {NoStop}%
\bibitem [{\citenamefont {Chang}\ \emph {et~al.}(2017)\citenamefont {Chang},
  \citenamefont {Xu}, \citenamefont {Wieder}, \citenamefont {Sanchez},
  \citenamefont {Huang}, \citenamefont {Belopolski}, \citenamefont {Chang},
  \citenamefont {Zhang}, \citenamefont {Bansil}, \citenamefont {Lin},\ and\
  \citenamefont {Hasan}}]{Chang_2017}%
  \BibitemOpen
  \bibfield  {author} {\bibinfo {author} {\bibfnamefont {G.}~\bibnamefont
  {Chang}}, \bibinfo {author} {\bibfnamefont {S.-Y.}\ \bibnamefont {Xu}},
  \bibinfo {author} {\bibfnamefont {B.~J.}\ \bibnamefont {Wieder}}, \bibinfo
  {author} {\bibfnamefont {D.~S.}\ \bibnamefont {Sanchez}}, \bibinfo {author}
  {\bibfnamefont {S.-M.}\ \bibnamefont {Huang}}, \bibinfo {author}
  {\bibfnamefont {I.}~\bibnamefont {Belopolski}}, \bibinfo {author}
  {\bibfnamefont {T.-R.}\ \bibnamefont {Chang}}, \bibinfo {author}
  {\bibfnamefont {S.}~\bibnamefont {Zhang}}, \bibinfo {author} {\bibfnamefont
  {A.}~\bibnamefont {Bansil}}, \bibinfo {author} {\bibfnamefont
  {H.}~\bibnamefont {Lin}}, \ and\ \bibinfo {author} {\bibfnamefont {M.~Z.}\
  \bibnamefont {Hasan}},\ }\href@noop {} {\bibfield  {journal} {\bibinfo
  {journal} {Phys. Rev. Lett.}\ }\textbf {\bibinfo {volume} {119}},\ \bibinfo
  {pages} {206401} (\bibinfo {year} {2017})}\BibitemShut {NoStop}%
\bibitem [{\citenamefont {Pshenay-Severin}\ \emph {et~al.}(2018)\citenamefont
  {Pshenay-Severin}, \citenamefont {Ivanov}, \citenamefont {Burkov},\ and\
  \citenamefont {Burkov}}]{Pshenay_2018}%
  \BibitemOpen
  \bibfield  {author} {\bibinfo {author} {\bibfnamefont {D.~A.}\ \bibnamefont
  {Pshenay-Severin}}, \bibinfo {author} {\bibfnamefont {Y.~V.}\ \bibnamefont
  {Ivanov}}, \bibinfo {author} {\bibfnamefont {A.~A.}\ \bibnamefont {Burkov}},
  \ and\ \bibinfo {author} {\bibfnamefont {A.~T.}\ \bibnamefont {Burkov}},\
  }\href {\doibase 10.1088/1361-648x/aab0ba} {\bibfield  {journal} {\bibinfo
  {journal} {J. Phys.: Condens. Matter}\ }\textbf {\bibinfo {volume} {30}},\
  \bibinfo {pages} {135501} (\bibinfo {year} {2018})}\BibitemShut {NoStop}%
\bibitem [{\citenamefont {Burkov}\ \emph {et~al.}(2017)\citenamefont {Burkov},
  \citenamefont {Novikov}, \citenamefont {Zaitsev},\ and\ \citenamefont
  {Reith}}]{Burkov_2017}%
  \BibitemOpen
  \bibfield  {author} {\bibinfo {author} {\bibfnamefont {A.~T.}\ \bibnamefont
  {Burkov}}, \bibinfo {author} {\bibfnamefont {S.~V.}\ \bibnamefont {Novikov}},
  \bibinfo {author} {\bibfnamefont {V.~K.}\ \bibnamefont {Zaitsev}}, \ and\
  \bibinfo {author} {\bibfnamefont {H.}~\bibnamefont {Reith}},\ }\href@noop {}
  {\bibfield  {journal} {\bibinfo  {journal} {Semiconductors}\ }\textbf
  {\bibinfo {volume} {51}},\ \bibinfo {pages} {689} (\bibinfo {year}
  {2017})}\BibitemShut {NoStop}%
\bibitem [{\citenamefont {Burkov}\ \emph {et~al.}(2019)\citenamefont {Burkov},
  \citenamefont {Ivanov}, \citenamefont {K.Nielsch}, \citenamefont {Novikov},
  \citenamefont {Perez}, \citenamefont {Pshenay-Severin}, \citenamefont
  {Reith}, \citenamefont {Schnatmann}, \citenamefont {Schierning},\ and\
  \citenamefont {Volkov}}]{Burkov_2019}%
  \BibitemOpen
  \bibfield  {author} {\bibinfo {author} {\bibfnamefont {A.}~\bibnamefont
  {Burkov}}, \bibinfo {author} {\bibfnamefont {Y.}~\bibnamefont {Ivanov}},
  \bibinfo {author} {\bibnamefont {K.Nielsch}}, \bibinfo {author}
  {\bibfnamefont {S.}~\bibnamefont {Novikov}}, \bibinfo {author} {\bibfnamefont
  {N.}~\bibnamefont {Perez}}, \bibinfo {author} {\bibfnamefont
  {D.}~\bibnamefont {Pshenay-Severin}}, \bibinfo {author} {\bibfnamefont
  {H.}~\bibnamefont {Reith}}, \bibinfo {author} {\bibfnamefont
  {L.}~\bibnamefont {Schnatmann}}, \bibinfo {author} {\bibfnamefont
  {G.}~\bibnamefont {Schierning}}, \ and\ \bibinfo {author} {\bibfnamefont
  {M.}~\bibnamefont {Volkov}},\ }\href@noop {} {\bibfield  {journal} {\bibinfo
  {journal} {Materials Today: Proceedings}\ }\textbf {\bibinfo {volume} {8}},\
  \bibinfo {pages} {540} (\bibinfo {year} {2019})}\BibitemShut {NoStop}%
\bibitem [{\citenamefont {Sanchez}\ \emph {et~al.}(2019)\citenamefont
  {Sanchez}, \citenamefont {Belopolski}, \citenamefont {Cochran}, \citenamefont
  {Xu}, \citenamefont {Yin}, \citenamefont {Chang}, \citenamefont {Xie},
  \citenamefont {Manna}, \citenamefont {S\"u\ss}, \citenamefont {Huang},
  \citenamefont {Alidoust}, \citenamefont {Multer}, \citenamefont {Zhang},
  \citenamefont {Shumiya}, \citenamefont {Wang}, \citenamefont {Wang},
  \citenamefont {Chang}, \citenamefont {Felser}, \citenamefont {Xu},
  \citenamefont {Jia}, \citenamefont {Lin},\ and\ \citenamefont
  {Hasan}}]{Sanchez_2019}%
  \BibitemOpen
  \bibfield  {author} {\bibinfo {author} {\bibfnamefont {D.~S.}\ \bibnamefont
  {Sanchez}}, \bibinfo {author} {\bibfnamefont {I.}~\bibnamefont {Belopolski}},
  \bibinfo {author} {\bibfnamefont {T.~A.}\ \bibnamefont {Cochran}}, \bibinfo
  {author} {\bibfnamefont {X.}~\bibnamefont {Xu}}, \bibinfo {author}
  {\bibfnamefont {J.~X.}\ \bibnamefont {Yin}}, \bibinfo {author} {\bibfnamefont
  {G.}~\bibnamefont {Chang}}, \bibinfo {author} {\bibfnamefont
  {W.}~\bibnamefont {Xie}}, \bibinfo {author} {\bibfnamefont {K.}~\bibnamefont
  {Manna}}, \bibinfo {author} {\bibfnamefont {V.}~\bibnamefont {S\"u\ss}},
  \bibinfo {author} {\bibfnamefont {C.~Y.}\ \bibnamefont {Huang}}, \bibinfo
  {author} {\bibfnamefont {N.}~\bibnamefont {Alidoust}}, \bibinfo {author}
  {\bibfnamefont {D.}~\bibnamefont {Multer}}, \bibinfo {author} {\bibfnamefont
  {S.~S.}\ \bibnamefont {Zhang}}, \bibinfo {author} {\bibfnamefont
  {N.}~\bibnamefont {Shumiya}}, \bibinfo {author} {\bibfnamefont
  {X.}~\bibnamefont {Wang}}, \bibinfo {author} {\bibfnamefont {G.~Q.}\
  \bibnamefont {Wang}}, \bibinfo {author} {\bibfnamefont {T.~R.}\ \bibnamefont
  {Chang}}, \bibinfo {author} {\bibfnamefont {C.}~\bibnamefont {Felser}},
  \bibinfo {author} {\bibfnamefont {S.~Y.}\ \bibnamefont {Xu}}, \bibinfo
  {author} {\bibfnamefont {S.}~\bibnamefont {Jia}}, \bibinfo {author}
  {\bibfnamefont {H.}~\bibnamefont {Lin}}, \ and\ \bibinfo {author}
  {\bibfnamefont {M.~Z.}\ \bibnamefont {Hasan}},\ }\href@noop {} {\bibfield
  {journal} {\bibinfo  {journal} {Nature}\ }\textbf {\bibinfo {volume} {567}},\
  \bibinfo {pages} {500} (\bibinfo {year} {2019})}\BibitemShut {NoStop}%
\bibitem [{\citenamefont {Takane}\ \emph {et~al.}(2019)\citenamefont {Takane},
  \citenamefont {Wang}, \citenamefont {Souma}, \citenamefont {Nakayama},
  \citenamefont {Nakamura}, \citenamefont {Oinuma}, \citenamefont {Nakata},
  \citenamefont {Iwasawa}, \citenamefont {Cacho}, \citenamefont {Kim},
  \citenamefont {Horiba}, \citenamefont {Kumigashira}, \citenamefont
  {Takahashi}, \citenamefont {Ando},\ and\ \citenamefont {Sato}}]{Takane_2019}%
  \BibitemOpen
  \bibfield  {author} {\bibinfo {author} {\bibfnamefont {D.}~\bibnamefont
  {Takane}}, \bibinfo {author} {\bibfnamefont {Z.}~\bibnamefont {Wang}},
  \bibinfo {author} {\bibfnamefont {S.}~\bibnamefont {Souma}}, \bibinfo
  {author} {\bibfnamefont {K.}~\bibnamefont {Nakayama}}, \bibinfo {author}
  {\bibfnamefont {T.}~\bibnamefont {Nakamura}}, \bibinfo {author}
  {\bibfnamefont {H.}~\bibnamefont {Oinuma}}, \bibinfo {author} {\bibfnamefont
  {Y.}~\bibnamefont {Nakata}}, \bibinfo {author} {\bibfnamefont
  {H.}~\bibnamefont {Iwasawa}}, \bibinfo {author} {\bibfnamefont
  {C.}~\bibnamefont {Cacho}}, \bibinfo {author} {\bibfnamefont
  {T.}~\bibnamefont {Kim}}, \bibinfo {author} {\bibfnamefont {K.}~\bibnamefont
  {Horiba}}, \bibinfo {author} {\bibfnamefont {H.}~\bibnamefont {Kumigashira}},
  \bibinfo {author} {\bibfnamefont {T.}~\bibnamefont {Takahashi}}, \bibinfo
  {author} {\bibfnamefont {Y.}~\bibnamefont {Ando}}, \ and\ \bibinfo {author}
  {\bibfnamefont {T.}~\bibnamefont {Sato}},\ }\href@noop {} {\bibfield
  {journal} {\bibinfo  {journal} {Phys. Rev. Lett.}\ }\textbf {\bibinfo
  {volume} {122}},\ \bibinfo {pages} {076402} (\bibinfo {year}
  {2019})}\BibitemShut {NoStop}%
\bibitem [{\citenamefont {Rao}\ \emph {et~al.}(2019)\citenamefont {Rao},
  \citenamefont {Li}, \citenamefont {Zhang}, \citenamefont {Tian},
  \citenamefont {Li}, \citenamefont {Fu}, \citenamefont {Tang}, \citenamefont
  {Wang}, \citenamefont {Li}, \citenamefont {Fan}, \citenamefont {Li},
  \citenamefont {Huang}, \citenamefont {Liu}, \citenamefont {Long},
  \citenamefont {Fang}, \citenamefont {Weng}, \citenamefont {Shi},
  \citenamefont {Lei}, \citenamefont {Sun}, \citenamefont {Qian},\ and\
  \citenamefont {Ding}}]{Rao_2019}%
  \BibitemOpen
  \bibfield  {author} {\bibinfo {author} {\bibfnamefont {Z.}~\bibnamefont
  {Rao}}, \bibinfo {author} {\bibfnamefont {H.}~\bibnamefont {Li}}, \bibinfo
  {author} {\bibfnamefont {T.}~\bibnamefont {Zhang}}, \bibinfo {author}
  {\bibfnamefont {S.}~\bibnamefont {Tian}}, \bibinfo {author} {\bibfnamefont
  {C.}~\bibnamefont {Li}}, \bibinfo {author} {\bibfnamefont {B.}~\bibnamefont
  {Fu}}, \bibinfo {author} {\bibfnamefont {C.}~\bibnamefont {Tang}}, \bibinfo
  {author} {\bibfnamefont {L.}~\bibnamefont {Wang}}, \bibinfo {author}
  {\bibfnamefont {Z.}~\bibnamefont {Li}}, \bibinfo {author} {\bibfnamefont
  {W.}~\bibnamefont {Fan}}, \bibinfo {author} {\bibfnamefont {J.}~\bibnamefont
  {Li}}, \bibinfo {author} {\bibfnamefont {Y.}~\bibnamefont {Huang}}, \bibinfo
  {author} {\bibfnamefont {Z.}~\bibnamefont {Liu}}, \bibinfo {author}
  {\bibfnamefont {Y.}~\bibnamefont {Long}}, \bibinfo {author} {\bibfnamefont
  {C.}~\bibnamefont {Fang}}, \bibinfo {author} {\bibfnamefont {H.}~\bibnamefont
  {Weng}}, \bibinfo {author} {\bibfnamefont {Y.}~\bibnamefont {Shi}}, \bibinfo
  {author} {\bibfnamefont {H.}~\bibnamefont {Lei}}, \bibinfo {author}
  {\bibfnamefont {Y.}~\bibnamefont {Sun}}, \bibinfo {author} {\bibfnamefont
  {T.}~\bibnamefont {Qian}}, \ and\ \bibinfo {author} {\bibfnamefont
  {H.}~\bibnamefont {Ding}},\ }\href@noop {} {\bibfield  {journal} {\bibinfo
  {journal} {Nature}\ }\textbf {\bibinfo {volume} {567}},\ \bibinfo {pages}
  {496} (\bibinfo {year} {2019})}\BibitemShut {NoStop}%
\bibitem [{\citenamefont {Dutta}\ and\ \citenamefont
  {Pandey}(2021)}]{Dutta_2021}%
  \BibitemOpen
  \bibfield  {author} {\bibinfo {author} {\bibfnamefont {P.}~\bibnamefont
  {Dutta}}\ and\ \bibinfo {author} {\bibfnamefont {S.~K.}\ \bibnamefont
  {Pandey}},\ }\href@noop {} {\bibfield  {journal} {\bibinfo  {journal} {Eur.
  Phys. J. B}\ }\textbf {\bibinfo {volume} {94}},\ \bibinfo {pages} {81}
  (\bibinfo {year} {2021})}\BibitemShut {NoStop}%
\bibitem [{\citenamefont {Li}\ \emph {et~al.}(2005)\citenamefont {Li},
  \citenamefont {Ren}, \citenamefont {Zhang}, \citenamefont {Ito},\ and\
  \citenamefont {Wu}}]{Li_2005}%
  \BibitemOpen
  \bibfield  {author} {\bibinfo {author} {\bibfnamefont {C.~C.}\ \bibnamefont
  {Li}}, \bibinfo {author} {\bibfnamefont {W.~L.}\ \bibnamefont {Ren}},
  \bibinfo {author} {\bibfnamefont {L.~T.}\ \bibnamefont {Zhang}}, \bibinfo
  {author} {\bibfnamefont {K.}~\bibnamefont {Ito}}, \ and\ \bibinfo {author}
  {\bibfnamefont {J.~S.}\ \bibnamefont {Wu}},\ }\href@noop {} {\bibfield
  {journal} {\bibinfo  {journal} {Journal of Applied Physics}\ }\textbf
  {\bibinfo {volume} {98}},\ \bibinfo {pages} {063706} (\bibinfo {year}
  {2005})}\BibitemShut {NoStop}%
\bibitem [{\citenamefont {Ren}\ \emph {et~al.}(2005)\citenamefont {Ren},
  \citenamefont {Li}, \citenamefont {Zhang}, \citenamefont {Ito},\ and\
  \citenamefont {Wu}}]{Ren_2005}%
  \BibitemOpen
  \bibfield  {author} {\bibinfo {author} {\bibfnamefont {W.}~\bibnamefont
  {Ren}}, \bibinfo {author} {\bibfnamefont {C.}~\bibnamefont {Li}}, \bibinfo
  {author} {\bibfnamefont {L.}~\bibnamefont {Zhang}}, \bibinfo {author}
  {\bibfnamefont {K.}~\bibnamefont {Ito}}, \ and\ \bibinfo {author}
  {\bibfnamefont {J.}~\bibnamefont {Wu}},\ }\href {\doibase
  https://doi.org/10.1016/j.jallcom.2004.09.036} {\bibfield  {journal}
  {\bibinfo  {journal} {Journal of Alloys and Compounds}\ }\textbf {\bibinfo
  {volume} {392}},\ \bibinfo {pages} {50} (\bibinfo {year} {2005})}\BibitemShut
  {NoStop}%
\bibitem [{\citenamefont {Longhin}\ \emph {et~al.}(2017)\citenamefont
  {Longhin}, \citenamefont {Rizza}, \citenamefont {Viennois},\ and\
  \citenamefont {Papet}}]{Longhin_2017}%
  \BibitemOpen
  \bibfield  {author} {\bibinfo {author} {\bibfnamefont {M.}~\bibnamefont
  {Longhin}}, \bibinfo {author} {\bibfnamefont {M.}~\bibnamefont {Rizza}},
  \bibinfo {author} {\bibfnamefont {R.}~\bibnamefont {Viennois}}, \ and\
  \bibinfo {author} {\bibfnamefont {P.}~\bibnamefont {Papet}},\ }\href@noop {}
  {\bibfield  {journal} {\bibinfo  {journal} {Intermetallics}\ }\textbf
  {\bibinfo {volume} {88}},\ \bibinfo {pages} {46} (\bibinfo {year}
  {2017})}\BibitemShut {NoStop}%
\bibitem [{\citenamefont {Sun}\ \emph {et~al.}(2013)\citenamefont {Sun},
  \citenamefont {Lu},\ and\ \citenamefont {Morelli}}]{Sun_2013}%
  \BibitemOpen
  \bibfield  {author} {\bibinfo {author} {\bibfnamefont {H.}~\bibnamefont
  {Sun}}, \bibinfo {author} {\bibfnamefont {X.}~\bibnamefont {Lu}}, \ and\
  \bibinfo {author} {\bibfnamefont {D.~T.}\ \bibnamefont {Morelli}},\
  }\href@noop {} {\bibfield  {journal} {\bibinfo  {journal} {Journal of
  Electronic Materials}\ }\textbf {\bibinfo {volume} {42}},\ \bibinfo {pages}
  {1352} (\bibinfo {year} {2013})}\BibitemShut {NoStop}%
\bibitem [{\citenamefont {Kuo}\ \emph {et~al.}(2005)\citenamefont {Kuo},
  \citenamefont {Sivakumar}, \citenamefont {Huang},\ and\ \citenamefont
  {Lue}}]{Kuo_2005}%
  \BibitemOpen
  \bibfield  {author} {\bibinfo {author} {\bibfnamefont {Y.~K.}\ \bibnamefont
  {Kuo}}, \bibinfo {author} {\bibfnamefont {K.~M.}\ \bibnamefont {Sivakumar}},
  \bibinfo {author} {\bibfnamefont {S.~J.}\ \bibnamefont {Huang}}, \ and\
  \bibinfo {author} {\bibfnamefont {C.~S.}\ \bibnamefont {Lue}},\ }\href@noop
  {} {\bibfield  {journal} {\bibinfo  {journal} {Journal of Applied Physics}\
  }\textbf {\bibinfo {volume} {98}},\ \bibinfo {pages} {123510} (\bibinfo
  {year} {2005})}\BibitemShut {NoStop}%
\bibitem [{\citenamefont {Yu}\ \emph {et~al.}(2020)\citenamefont {Yu},
  \citenamefont {Kuang}, \citenamefont {Long}, \citenamefont {Ke},
  \citenamefont {Duan},\ and\ \citenamefont {Liu}}]{Jian_2020}%
  \BibitemOpen
  \bibfield  {author} {\bibinfo {author} {\bibfnamefont {J.}~\bibnamefont
  {Yu}}, \bibinfo {author} {\bibfnamefont {J.}~\bibnamefont {Kuang}}, \bibinfo
  {author} {\bibfnamefont {J.}~\bibnamefont {Long}}, \bibinfo {author}
  {\bibfnamefont {X.}~\bibnamefont {Ke}}, \bibinfo {author} {\bibfnamefont
  {X.}~\bibnamefont {Duan}}, \ and\ \bibinfo {author} {\bibfnamefont
  {Z.}~\bibnamefont {Liu}},\ }\href@noop {} {\bibfield  {journal} {\bibinfo
  {journal} {Journal of Materials Science: Materials in Electronics}\ }\textbf
  {\bibinfo {volume} {31}},\ \bibinfo {pages} {2139} (\bibinfo {year}
  {2020})}\BibitemShut {NoStop}%
\bibitem [{\citenamefont {Asanabe}(1965)}]{Asanabe_1965}%
  \BibitemOpen
  \bibfield  {author} {\bibinfo {author} {\bibfnamefont {S.}~\bibnamefont
  {Asanabe}},\ }\href@noop {} {\bibfield  {journal} {\bibinfo  {journal} {J.
  Phys. Soc. Jpn.}\ }\textbf {\bibinfo {volume} {20}},\ \bibinfo {pages} {933}
  (\bibinfo {year} {1965})}\BibitemShut {NoStop}%
\bibitem [{\citenamefont {Dutta}\ and\ \citenamefont
  {Pandey}(2018)}]{Dutta_2018}%
  \BibitemOpen
  \bibfield  {author} {\bibinfo {author} {\bibfnamefont {P.}~\bibnamefont
  {Dutta}}\ and\ \bibinfo {author} {\bibfnamefont {S.~K.}\ \bibnamefont
  {Pandey}},\ }\href@noop {} {\bibfield  {journal} {\bibinfo  {journal}
  {Comput. Condens. Matter}\ }\textbf {\bibinfo {volume} {16}},\ \bibinfo
  {pages} {e00325} (\bibinfo {year} {2018})}\BibitemShut {NoStop}%
\bibitem [{\citenamefont {Dutta}\ and\ \citenamefont
  {Pandey}(2019)}]{Dutta_2019}%
  \BibitemOpen
  \bibfield  {author} {\bibinfo {author} {\bibfnamefont {P.}~\bibnamefont
  {Dutta}}\ and\ \bibinfo {author} {\bibfnamefont {S.~K.}\ \bibnamefont
  {Pandey}},\ }\href@noop {} {\bibfield  {journal} {\bibinfo  {journal} {J.
  Phys. Condens. Matter}\ }\textbf {\bibinfo {volume} {31}},\ \bibinfo {pages}
  {145602} (\bibinfo {year} {2019})}\BibitemShut {NoStop}%
\bibitem [{\citenamefont {Blaha}\ \emph {et~al.}(2001)\citenamefont {Blaha},
  \citenamefont {Schwarz}, \citenamefont {Madsen}, \citenamefont {Kvasnicka},\
  and\ \citenamefont {Luitz}}]{Blaha_2001}%
  \BibitemOpen
  \bibfield  {author} {\bibinfo {author} {\bibfnamefont {P.}~\bibnamefont
  {Blaha}}, \bibinfo {author} {\bibfnamefont {K.}~\bibnamefont {Schwarz}},
  \bibinfo {author} {\bibfnamefont {G.~K.~H.}\ \bibnamefont {Madsen}}, \bibinfo
  {author} {\bibfnamefont {D.}~\bibnamefont {Kvasnicka}}, \ and\ \bibinfo
  {author} {\bibfnamefont {J.}~\bibnamefont {Luitz}},\ }\href@noop {}
  {\bibfield  {journal} {\bibinfo  {journal} {An augmented plane wave + local
  orbitals program for calculating crystal properties}\ } (\bibinfo {year}
  {2001})}\BibitemShut {NoStop}%
\bibitem [{\citenamefont {Bo\'{r}en}\ and\ \citenamefont
  {Kemi}(1933)}]{Boren_1933}%
  \BibitemOpen
  \bibfield  {author} {\bibinfo {author} {\bibfnamefont {B.}~\bibnamefont
  {Bo\'{r}en}}\ and\ \bibinfo {author} {\bibfnamefont {A.}~\bibnamefont
  {Kemi}},\ }\href@noop {} {\bibfield  {journal} {\bibinfo  {journal} {Min.
  Geol.}\ }\textbf {\bibinfo {volume} {11A}},\ \bibinfo {pages} {1} (\bibinfo
  {year} {1933})}\BibitemShut {NoStop}%
\bibitem [{\citenamefont {Perdew}\ \emph {et~al.}(2008)\citenamefont {Perdew},
  \citenamefont {Ruzsinszky}, \citenamefont {Csonka}, \citenamefont {Vydrov},
  \citenamefont {Scuseria}, \citenamefont {Constantin}, \citenamefont {Zhou},\
  and\ \citenamefont {Burke}}]{Perdew_2008}%
  \BibitemOpen
  \bibfield  {author} {\bibinfo {author} {\bibfnamefont {J.~P.}\ \bibnamefont
  {Perdew}}, \bibinfo {author} {\bibfnamefont {A.}~\bibnamefont {Ruzsinszky}},
  \bibinfo {author} {\bibfnamefont {G.~I.}\ \bibnamefont {Csonka}}, \bibinfo
  {author} {\bibfnamefont {O.~A.}\ \bibnamefont {Vydrov}}, \bibinfo {author}
  {\bibfnamefont {G.~E.}\ \bibnamefont {Scuseria}}, \bibinfo {author}
  {\bibfnamefont {L.~A.}\ \bibnamefont {Constantin}}, \bibinfo {author}
  {\bibfnamefont {X.}~\bibnamefont {Zhou}}, \ and\ \bibinfo {author}
  {\bibfnamefont {K.}~\bibnamefont {Burke}},\ }\href {\doibase
  10.1103/PhysRevLett.100.136406} {\bibfield  {journal} {\bibinfo  {journal}
  {Phys. Rev. Lett.}\ }\textbf {\bibinfo {volume} {100}},\ \bibinfo {pages}
  {136406} (\bibinfo {year} {2008})}\BibitemShut {NoStop}%
\bibitem [{\citenamefont {Madsen}\ and\ \citenamefont
  {Singh}(2006)}]{Madsen_2006}%
  \BibitemOpen
  \bibfield  {author} {\bibinfo {author} {\bibfnamefont {G.~K.}\ \bibnamefont
  {Madsen}}\ and\ \bibinfo {author} {\bibfnamefont {D.~J.}\ \bibnamefont
  {Singh}},\ }\href@noop {} {\bibfield  {journal} {\bibinfo  {journal} {Comput.
  Phys. Commun.}\ }\textbf {\bibinfo {volume} {175}},\ \bibinfo {pages} {67}
  (\bibinfo {year} {2006})}\BibitemShut {NoStop}%
\bibitem [{\citenamefont {Ashcroft}\ and\ \citenamefont
  {Mermin}(2011)}]{ashcroft}%
  \BibitemOpen
  \bibfield  {author} {\bibinfo {author} {\bibfnamefont {N.}~\bibnamefont
  {Ashcroft}}\ and\ \bibinfo {author} {\bibfnamefont {N.}~\bibnamefont
  {Mermin}},\ }\href@noop {} {\emph {\bibinfo {title} {Solid State Physics}}}\
  (\bibinfo  {publisher} {Cengage Learning},\ \bibinfo {year}
  {2011})\BibitemShut {NoStop}%
\bibitem [{\citenamefont {Antonov}\ \emph {et~al.}(2019)\citenamefont
  {Antonov}, \citenamefont {Ivanov}, \citenamefont {Konstantinov},
  \citenamefont {Kuznetsova}, \citenamefont {Novikov}, \citenamefont
  {Ovchinnikov}, \citenamefont {Pshenay-Severin},\ and\ \citenamefont
  {Burkov}}]{Antonov_2019}%
  \BibitemOpen
  \bibfield  {author} {\bibinfo {author} {\bibfnamefont {A.}~\bibnamefont
  {Antonov}}, \bibinfo {author} {\bibfnamefont {Y.}~\bibnamefont {Ivanov}},
  \bibinfo {author} {\bibfnamefont {P.}~\bibnamefont {Konstantinov}}, \bibinfo
  {author} {\bibfnamefont {V.}~\bibnamefont {Kuznetsova}}, \bibinfo {author}
  {\bibfnamefont {S.}~\bibnamefont {Novikov}}, \bibinfo {author} {\bibfnamefont
  {A.}~\bibnamefont {Ovchinnikov}}, \bibinfo {author} {\bibfnamefont
  {D.}~\bibnamefont {Pshenay-Severin}}, \ and\ \bibinfo {author} {\bibfnamefont
  {A.}~\bibnamefont {Burkov}},\ }\href@noop {} {\bibfield  {journal} {\bibinfo
  {journal} {Journal of Applied Physics}\ }\textbf {\bibinfo {volume} {126}},\
  \bibinfo {pages} {245103} (\bibinfo {year} {2019})}\BibitemShut {NoStop}%
\bibitem [{\citenamefont {Ovchinnikov}\ \emph {et~al.}(2019)\citenamefont
  {Ovchinnikov}, \citenamefont {Konstantinov}, \citenamefont
  {Pshenay-Severin},\ and\ \citenamefont {Burkov}}]{Ovchinnikov_2019}%
  \BibitemOpen
  \bibfield  {author} {\bibinfo {author} {\bibfnamefont {A.~Y.}\ \bibnamefont
  {Ovchinnikov}}, \bibinfo {author} {\bibfnamefont {P.~P.}\ \bibnamefont
  {Konstantinov}}, \bibinfo {author} {\bibfnamefont {D.~A.}\ \bibnamefont
  {Pshenay-Severin}}, \ and\ \bibinfo {author} {\bibfnamefont {A.~T.}\
  \bibnamefont {Burkov}},\ }\href@noop {} {\bibfield  {journal} {\bibinfo
  {journal} {Semiconductors}\ }\textbf {\bibinfo {volume} {53}},\ \bibinfo
  {pages} {737} (\bibinfo {year} {2019})}\BibitemShut {NoStop}%
\end{thebibliography}%
\bibliographystyle{apsrev4-1}

\end{document}